\pdfoutput=1
\documentclass[11pt,twoside,a4paper,cmspaper,final,collab]{cms-tdr}
\def\svnVersion{a9cfb18}\def\svnDate{2023/06/19}\def\cmsCernNoTag{CERN-EP-2022-051}\def\cmsCernDate{\today}\def\cmsMessage{Published in the Journal of High Energy Physics as \href{http://dx.doi.org/10.1007/JHEP06(2023)081}{\doi{10.1007/JHEP06(2023)081}.}}
\begin{document}\cmsNoteHeader{TOP-20-005}

\newcommand{\Acp}{\ensuremath{A_\text{CP}}\xspace}
\newcommand{\Acpprime}{\ensuremath{A'_\text{CP}}\xspace}
\newcommand{\Mlb}{\ensuremath{m_{\Pell\PQb}}\xspace}
\newcommand{\MW}{\ensuremath{m_{\PW}}\xspace}
\newcommand{\Mt}{\ensuremath{m_{\PQt}}\xspace}
\newcommand{\Mtcub}{\ensuremath{m_{\PQt}^3}\xspace}
\newcommand{\pp}{\ensuremath{\Pp\Pp}\xspace}
\newcommand{\Wjets}{\ensuremath{\PW\text{+jets}}\xspace}
\newcommand{\ejets}{\ensuremath{\Pe\text{+jets}}\xspace}
\newcommand{\mjets}{\ensuremath{\PGm\text{+jets}}\xspace}
\newcommand{\WHF}{\ensuremath{\PW\text{+HF}}\xspace}
\newcommand{\dtG}{\ensuremath{d_{\PQt\mathrm{G}}}\xspace}
\newcommand{\dtg}{\ensuremath{d_\PQt^\Pg}\xspace}
\newcommand{\dt}{\ensuremath{\hat{d}_\PQt}\xspace}
\newcommand{\newTeV}{\ensuremath{\sqrt{s}=13\TeV}\xspace}
\newcommand{\oldTeV}{\ensuremath{\sqrt{s}=8\TeV}\xspace}
\newcommand{\SCC}{\ensuremath{g_s}\xspace}
\newcommand{\SUTG}{\ensuremath{T^a}\xspace}
\newcommand{\sigmn}{\ensuremath{\sigma^{\mu\nu}}\xspace}
\newcommand{\CMDM}{\ensuremath{a_\PQt^\Pg}\xspace}
\newcommand{\GFST}{\ensuremath{G_{\mu\nu}^a}\xspace}
\newcommand{\Othree}{\ensuremath{O_3}\xspace}
\newcommand{\Osix}{\ensuremath{O_6}\xspace}
\newcommand{\Otwelve}{\ensuremath{O_{12}}\xspace}
\newcommand{\Ofourteen}{\ensuremath{O_{14}}\xspace}
\newcommand{\Qell}{\ensuremath{Q_{\Pell}}\xspace}
\newcommand{\momB}{\ensuremath{p_{\PQb}}\xspace}
\newcommand{\momAB}{\ensuremath{p_{\PAQb}}\xspace}
\newcommand{\momL}{\ensuremath{p_{\Pell}}\xspace}
\newcommand{\momJ}{\ensuremath{p_{j_1}}\xspace}
\newcommand{\vecB}{\ensuremath{\vec{p}_{\PQb}}\xspace}
\newcommand{\vecAB}{\ensuremath{\vec{p}_{\PAQb}}\xspace}
\newcommand{\vecL}{\ensuremath{\vec{p}_{\Pell}}\xspace}
\newcommand{\vecJ}{\ensuremath{\vec{p}_{j_1}}\xspace}
\newcommand{\epsLC}{\ensuremath{\epsilon_{\mu\nu\alpha\beta}}\xspace}
\newcommand{\abseta}{\ensuremath{\abs{\eta}}\xspace}
\newcommand{\Irel}{\ensuremath{I_{\text{rel}}}\xspace}
\newcommand{\DeepCSV}{\ensuremath{\textsc{DeepCSV}}\xspace}
\newcommand{\chisq}{\ensuremath{\chi^2}\xspace}
\newcommand{\mjjb}{\ensuremath{m_{\mathrm{jj}\PQb}}\xspace}
\newcommand{\mjj}{\ensuremath{m_{\mathrm{jj}}}\xspace}
\newcommand{\sigt}{\ensuremath{\sigma_{\PQt}}\xspace}
\newcommand{\sigW}{\ensuremath{\sigma_{\PW}}\xspace}
\newcommand{\Oi}{\ensuremath{O_i}\xspace}
\newcommand{\Npos}{\ensuremath{N_{\text{events}}(\Oi>0)}\xspace}
\newcommand{\Nneg}{\ensuremath{N_{\text{events}}(\Oi<0)}\xspace}
\newcommand{\numb}{\ensuremath{n_b}\xspace}
\newcommand{\numpos}{\ensuremath{n_s^+}\xspace}
\newcommand{\numneg}{\ensuremath{n_s^-}\xspace}
\newcommand{\Numpos}{\ensuremath{N^+}\xspace}
\newcommand{\Numneg}{\ensuremath{N^-}\xspace}
\newcommand{\ratiob}{\ensuremath{r_b}\xspace}
\newcommand{\Mlbi}{\ensuremath{\Mlb^i}\xspace}
\newcommand{\fspos}{\ensuremath{f_s^+}\xspace}
\newcommand{\fsneg}{\ensuremath{f_s^-}\xspace}
\newcommand{\fbpos}{\ensuremath{f_b^+}\xspace}
\newcommand{\fbneg}{\ensuremath{f_b^-}\xspace}
\newcommand{\epsc}{\ensuremath{\epsilon_c}\xspace}
\newcommand{\epsw}{\ensuremath{\epsilon_w}\xspace}
\newcommand{\dilution}{\ensuremath{D}\xspace}
\newcommand{\delvec}{\ensuremath{\begin{pmatrix}
    \frac{\partial\dtG}{\partial \Acp} \\[5pt]
    \frac{\partial\dtG}{\partial a} \\[5pt]
    \frac{\partial\dtG}{\partial b} \\[5pt]
    \frac{\partial\dtG}{\partial c} \\[5pt]
    \frac{\partial\dtG}{\partial d}
\end{pmatrix}}}
\newcommand{\DsqdtG}{\ensuremath{\Delta^2_{\dtG}}\xspace}
\newcommand{\DsqAcp}{\ensuremath{\Delta^2_{\Acp}}\xspace}
\newcommand{\Dsqa}{\ensuremath{\Delta^2_{a}}\xspace}
\newcommand{\Dsqb}{\ensuremath{\Delta^2_{b}}\xspace}
\newcommand{\Dsqc}{\ensuremath{\Delta^2_{c}}\xspace}
\newcommand{\Dsqd}{\ensuremath{\Delta^2_{d}}\xspace}
\newlength\cmsTabSkip\setlength{\cmsTabSkip}{2ex}

\cmsNoteHeader{TOP-20-005}
\title{Search for CP violation using \texorpdfstring{\ttbar}{ttbar} events in the lepton+jets channel in \texorpdfstring{\pp}{pp} collisions at \texorpdfstring{\newTeV}{sqrt(s) = 13 TeV}}
\date{\today}

\abstract{Results are presented on a search for CP violation in the production and decay of top quark-antiquark pairs in the lepton+jets channel. The search is based on data from proton-proton collisions at \newTeV, collected with the CMS detector, corresponding to an integrated luminosity of 138\fbinv. Possible CP violation effects are evaluated by measuring asymmetries in observables constructed from linearly independent four-momentum vectors of the final-state particles. The dimensionless chromoelectric dipole moment of the top quark obtained from the observed asymmetries is measured to be $0.04\pm 0.10\stat\pm 0.07\syst$, and the asymmetries exhibit no evidence for CP-violating effects, consistent with expectations from the standard model.}

\hypersetup{%
pdfauthor={CMS Collaboration},%
pdftitle={Search for CP violation using ttbar events in the lepton+jets channel in pp collisions at sqrt(s) = 13 TeV},%
pdfsubject={CMS},%
pdfkeywords={CMS, top quarks}}

\maketitle

\section{Introduction}
The standard model (SM) of particle physics predicts the violation of the combined charge conjugation and parity (CP) symmetry that originates from a complex phase in the Cabibbo--Kobayashi--Maskawa matrix~\cite{CPVtop:CKMmatrix1973,CPVtop:CKMmatrix}.
Measurements of CP violation (CPV) in the strange (\PQs), bottom (\PQb), and charm (\PQc) quark sectors conducted over the past few decades~\cite{NA48:1999szy,CPVtop:Bfactories,Pajero:2020dum} have been found to be consistent with the SM expectations.
However, the level of CPV in the SM is insufficient to accommodate the observed matter-antimatter asymmetry in the universe~\cite{CPVtop:RevParticlePhy}, motivating searches for sources of CPV beyond the SM (BSM).
In contrast to the \PQs, \PQc, and \PQb quark sectors, CPV in the top (\PQt) quark sector is relatively unexplored.
In the SM, the CPV effects in top quark pair (\ttbar) decays are expected to be small due to the large mass of the top quark in comparison with the other quarks, leading to the Glashow--Iliopoulos--Maiani cancellation~\cite{Glashow:1970gm}.
Thus, any observed CP-violating asymmetry would indicate the presence of BSM phenomena~\cite{CPVtop:CPVinTOP}.
For example, a nonzero chromoelectric dipole moment (CEDM) of the top quark~\cite{CPVtop:8TeVRef,CPVtop:13TeVRef,CPVtop:Tevatron,CPVtop:CPVsource} can generate sizable CPV in the production of \ttbar.
Previous studies performed by CMS in data from \pp collisions at \oldTeV~\cite{CPVtop:CMSresult} found the CP-violating asymmetries (\Acp) in the \ttbar lepton+jets channel to be consistent with the SM prediction.

This paper presents the results of new searches by the CMS Collaboration for CP-violating asymmetries in \ttbar events using the lepton+jets channel from \pp collisions produced at the LHC.
A possible source of CPV at the top quark production and decay vertices arises from BSM interactions through the CEDM of the top quark.
In a model with contributions from a CEDM~\cite{CPVtop:13TeVRef}, the magnetic and electric couplings between top quarks and gluons (\Pg) are conventionally written as
\begin{linenomath}\begin{equation}\label{eq:lagrangian}
    \mathcal{L}=\frac{\SCC}{2}\PAQt\SUTG\sigmn(\CMDM+i\gamma_5\dtg)\PQt\GFST,
\end{equation}\end{linenomath}
where \SCC and \GFST are the strong coupling constant and the gluon field strength tensor, respectively; \PQt and \PAQt are the wavefunctions of the top quark and antiquark; \SUTG are $SU(3)$ generators; \sigmn is defined by the operator $\frac{i}{2}[\gamma^\mu, \gamma^\nu]$; \CMDM refers to the parameter of the chromomagnetic dipole moment; and \dtg is the CP-odd CEDM.
From Ref.~\cite{CPVtop:13TeVRef}, \dtg can be converted into a dimensionless CEDM parameter \dtG as
\begin{linenomath}\begin{equation}\label{eq:cedm_conversion}
    \dtg = \frac{\sqrt{2}v}{\Lambda^2}\text{Im}(\dtG),
\end{equation}\end{linenomath}
where $\Lambda$ is a high-mass scale of the BSM phenomena and $v$ is the vacuum expectation value for the Higgs boson field ($v \approx 246\GeV$).
Higher \dtG values are expected to yield larger \Acp contributions.

In the lepton+jets channel, one of the top quarks is presumed to decay into a bottom quark and a \PW boson that subsequently decays into quark pairs (\qqbar).
The other top quark is required to decay into a bottom quark and a \PW boson that decays leptonically into an electron or muon and its associated neutrino.
We will refer to this as the leptonically decaying top quark.
The analysis exploits four T-odd observables, where T is the time-reversal operator, as proposed in Ref.~\cite{CPVtop:13TeVRef}.
The CP observables are chosen to come from reconstructable final-state objects that can be well measured.
For example, some observables have been discarded because they need the momentum of the leptonically decaying top quark, which is not experimentally measured.
The CP observables take the form $\vec{v_1} \cdot (\vec{v_2}\times\vec{v_3})$, where $\vec{v_i}$ are spin or momentum vectors and $i=1\text{--}3$~\cite{CPVtop:8TeVRef,CPVtop:13TeVRef,CPVtop:Tevatron}.
These triple-product observables are odd under CP transformation if CPT is conserved.
The four CP observables measured in this analysis are defined as
\begin{linenomath}\begin{equation}\begin{aligned}
    \Othree &= \Qell\epsilon(\momB,\,\momAB,\,\momL,\,\momJ) \propto \Qell\vecB^\ast\cdot(\vecL^\ast\times\vecJ^\ast),\\
    \Osix &= \Qell\epsilon(P,\,\momB-\momAB,\,\momL,\,\momJ) \propto
    \Qell(\vecB-\vecAB) \cdot (\vecL\times\vecJ),\\
    \Otwelve &= q\cdot (\momB-\momAB)\epsilon(P,\,q,\,\momB,\,\momAB) \propto (\vecB-\vecAB)_z \cdot (\vecB\times\vecAB)_z,\\
    \Ofourteen &= \epsilon(P,\,\momB+\momAB,\,\momL,\,\momJ) \propto (\vecB+\vecAB)\cdot(\vecL\times\vecJ).
\end{aligned}\end{equation}\end{linenomath}
The symbol $\propto$ indicates that the CP observable is proportional to the triple product; the asterisk symbol represents the quantity measured in the center-of-mass frame of the $\bbbar$ pair, where \PAQb indicates the \PQb antiquark; $\epsilon(a,b,c,d) \equiv \epsLC a^\mu b^\nu c^\alpha d^\beta$, where \epsLC is the Levi--Civita tensor; $P$ and $q$ are the sum and difference of the four-momenta of the protons in the \pp collision, respectively; \momB and \momAB refer to the two \PQb jet momenta, where the \PQb jet definition will be given below; \momL is the momentum of the lepton (\Pell) that originates from the \PW boson decay; \momJ refers to the momentum of the highest transverse momentum (\pt) jet from the hadronically decaying \PW boson; \Qell is the charge of the lepton; and the $z$ subscript indicates a projection along the beam axis in the CMS coordinate system.

The tabulated results are provided in the HEPData record for this analysis~\cite{hepdata}.
The paper is organized as follows.
Section~\ref{sec:detector} introduces the basic features of the CMS detector.
Sections~\ref{sec:sample} and~\ref{sec:selection} provide information on the data, simulations, and selection criteria.
Sections~\ref{sec:fitresult} and~\ref{sec:uncertainty} describe the fitting procedures, the instrumental effects, and the resulting systematic uncertainties.
The final results are presented in Section~\ref{sec:results}, with a brief summary given in Section~\ref{sec:summary}.

\section{The CMS detector}\label{sec:detector}
The central feature of the CMS apparatus is a superconducting solenoid of 6\unit{m} internal diameter, providing a magnetic field of 3.8\unit{T}.
Within the solenoid volume are a silicon pixel and strip tracker, a lead tungstate crystal electromagnetic calorimeter (ECAL), and a brass and scintillator hadron calorimeter (HCAL), each composed of a barrel and two endcap sections.
Forward calorimeters extend the pseudorapidity ($\eta$) coverage provided by the barrel and endcap detectors. Muons are detected in gas-ionization chambers embedded in the steel flux-return yoke outside the solenoid.

Events of interest are selected using a two-tiered trigger system~\cite{Khachatryan:2016bia}.
The first level is composed of specialized hardware processors using information from the calorimeters and muon detectors to select events at a rate of around 100\unit{kHz}.
The second level, known as the high-level trigger, consists of a farm of processors running a version of the full event reconstruction software optimized for fast processing, which reduces the event rate to around 1\unit{kHz} before data storage~\cite{CMS:HLT}.

A more detailed description of the CMS detector, together with a definition of the coordinate system and kinematic variables, can be found in Ref.~\cite{CMS:detector}.

\section{Data and simulated samples}\label{sec:sample}
This study involves data from \pp collisions at \newTeV collected with the CMS detector in 2016--2018, corresponding to an integrated luminosity of 138\fbinv with an overall luminosity uncertainty of 1.6\%~\cite{CMS:lumi2016, CMS:lumi2017, CMS:lumi2018}.

The \ttbar production events are simulated using Monte Carlo (MC) programs.
The \ttbar is simulated using quantum chromodynamics (QCD) at next-to-leading-order (NLO) precision through the matrix element (ME) in the \POWHEG 2.0 event generator~\cite{Sim:powheg1, Sim:powheg2, Sim:powheg3, Sim:powheg4}.
The value of the top quark mass (\Mt) is set to 172.5\GeV.
The \POWHEG output is combined with the parton shower (PS) simulation of \PYTHIA 8.205~\cite{Sim:pythia2}, with the underlying-event (UE) tune CP5~\cite{Sim:CP5}.
The parton distribution functions (PDFs) NNPDF 3.1~\cite{Sim:NNPDF3.1} at next-to-NLO order (NNLO) are used to model the data.
Samples of \ttbar with different values of CEDM are simulated with the \MADGRAPH generator~\cite{Sim:madgraph} at leading-order (LO) precision interfaced with \PYTHIA.
These samples are used as a cross-check of the model dependency of each CP observable.

Several backgrounds are considered with single top quark production being the leading contribution.
This is simulated at NLO using \MADGRAPH 2.4.2 with the FxFx matching scheme~\cite{Sim:FXFX} for $s$-channel production and \POWHEG for $t$-channel and $\PQt\PW$ production.
All samples are interfaced with \PYTHIA through the UE tune CUETP8M1~\cite{Sim:CUETP8M1} for $t$-channel production for the 2016 data and CP5 for the rest.
Diboson ($\PV\PV$), \Wjets, Drell--Yan (DY), and QCD multijet productions are simulated at LO using the \MADGRAPH 2.4.4 generator.
They are then interfaced with \PYTHIA using the MLM matching scheme~\cite{Sim:MLMmatching}.
Events coming from the decay of \ttbar into either dilepton+jets or hadronic multijets can be incorrectly included in the signal event sample due to particle misidentification.
They are considered as a background.
The background from \PW boson events with heavy-flavor quarks (\WHF), which is not a large background, is important during the estimation of the systematic uncertainties, and will be discussed in detail in Section~\ref{sec:uncertainty}.

The simulation of the experimental apparatus is based on \GEANTfour~\cite{Sim:geant4}.
To model the effect of additional \pp interactions within the same or nearby bunch crossings (pileup), simulated minimum-bias interactions are included in the simulated samples~\cite{Sim:pileup}.
The number of pileup interactions in the simulation is reweighted to match the distribution in data.

\section{Physical objects and event selection}\label{sec:selection}
The event selection criteria are based on the lepton+jets channel signature of the \ttbar process.
The signal events are required to contain one reconstructed lepton, and at least four reconstructed jets, including two \PQb-tagged jets (defined below) originating from \ttbar decays.
The events are further categorized into two channels based on the lepton flavor: either electron or muon.
Events are selected at the trigger level using single-lepton triggers, which require the presence of an isolated lepton above a \PT threshold in the range 24--35\GeV, depending on the year, and $\abseta < 2.4$.
The same trigger selection is applied to data and to the simulated samples.
All physics objects are reconstructed offline using a particle-flow (PF) algorithm~\cite{CMS:particle_flow}.
The aim of the algorithm is to reconstruct all final-state particles (photons, charged and neutral hadrons, muons, and electrons) in the event, using combined information from the CMS detector.
The reconstructed vertex with the largest sum of object $\PT^2$ is taken to be the primary \pp interaction vertex.

Electron candidates are identified using the combination of the silicon tracker and the corresponding ECAL cluster information~\cite{CMS:el_PF}.
Electron energies are determined using the momenta derived from the electron track in the tracker system, the energy of the spatially compatible ECAL cluster, and the sum of all compatible bremsstrahlung photon energies.
In order to reject events with misidentified electron candidates and candidates originating from photon conversions, additional electron identification requirements are applied.
A PF-based combined relative isolation, \Irel, is defined as the \PT sum of all neutral hadron, charged hadron, and photon candidates within a cone of size $\DR = \sqrt{\smash[b]{(\Delta\eta)^2+(\Delta\phi)^2}} < 0.4$ (where $\phi$ is the azimuthal angle in radians) around the lepton direction, divided by the lepton \PT, with a correction to suppress the residual effect of pileup~\cite{CMS:el_reliso}.
The isolated electron candidate is required to have $\PT>38\GeV$, $\abseta < 2.4$, and $\Irel < 0.0287+0.506/\PT$ ($0.0445+0.963/\PT$) in the barrel (endcaps), with \PT in \GeV.
Due to reduced reconstruction efficiency, the gap ($1.44 < \abseta < 1.57$) between the barrel and endcap parts of the ECAL is excluded.
Events with additional electrons satisfying a looser set of selection criteria with a lower-\PT threshold and a looser isolation requirement are rejected to reduce the background contribution (\eg, from DY production).

Muon candidates are reconstructed using information obtained in the tracker, combined with the muon system~\cite{CMS:mu_PF}.
Identification methods are applied to reject muon candidates that are misidentified or muons that originated from decay-in-flight.
The isolated muon candidate is required to have $\PT > 30\GeV$, $\abseta < 2.4$, and $\Irel < 0.15$.
Events with additional muons satisfying a looser set of selection criteria are rejected as well.

Jets are clustered using the anti-\kt algorithm with a distance parameter of 0.4~\cite{CMS:fastjet, CMS:antikt}.
The jet \ptvec is defined as the vectorial sum of the momenta of all PF candidates in the jet anti-\kt clustering.
Pileup can contribute additional tracks and calorimetric energy depositions to the jet momentum.
To mitigate this effect, tracks identified as originating from pileup vertices are discarded, and an offset correction is applied to correct for remaining contributions from neutral particles from pileup~\cite{CMS:particle_flow}.
Corrections to the jet energy are applied as a function of jet \PT and $\eta$ by studying discrepancies between simulation and data.
At least four jets are required with $\PT > 30\GeV$, $\abseta < 2.4$, and angular separation relative to the selected lepton of $\DR > 0.4$.
Jets are identified as arising from the hadronization of \PQb quarks, denoted as either \PQb-tagged jets or \PQb jets, using the deep-learned combined secondary-vertex algorithm (\DeepCSV).
This combines the information for a given jet from track impact parameters and secondary vertices into a model optimized through a deep neural network.
The selected working point has a signal identification efficiency of 68\%, with a probability to misidentify \PQc quark, and light-flavor quark and gluon jets as \PQb jets of approximately 12 and 1.1\% in \ttbar events, respectively~\cite{CMS:bjet_eff}.

The top quark and antiquark candidates associated with \PW bosons decaying to \qqbar are reconstructed using one of the \PQb-tagged jets and two non-\PQb-tagged jets in the event through a \chisq algorithm that uses the top quark and \PW boson masses as constraints~\cite{CMS:sorting_algo}.
Those candidates are chosen from the combination having the lowest \chisq value, with the \chisq defined as
\begin{linenomath}\begin{equation}
    \chisq = \left(\frac{\mjjb-\Mt}{\sigt}\right)^2 + \left(\frac{\mjj-\MW}{\sigW}\right)^2,
\end{equation}\end{linenomath}
where \mjjb is the invariant mass of the two non-\PQb-tagged jets and the associated \PQb-tagged jet; \Mt and \sigt are the default top quark mass and average top quark invariant mass resolution of 172.5 and 16.3\GeV, respectively; \mjj is the invariant mass of the two non-\PQb-tagged jets; and \MW and \sigW are the default \PW boson mass and average \PW boson invariant mass resolution of 82.9 and 9.5\GeV, respectively~\cite{CPVtop:chi2value}.

\begin{figure}[!p]
    \centering
    \includegraphics[width=0.45\textwidth]{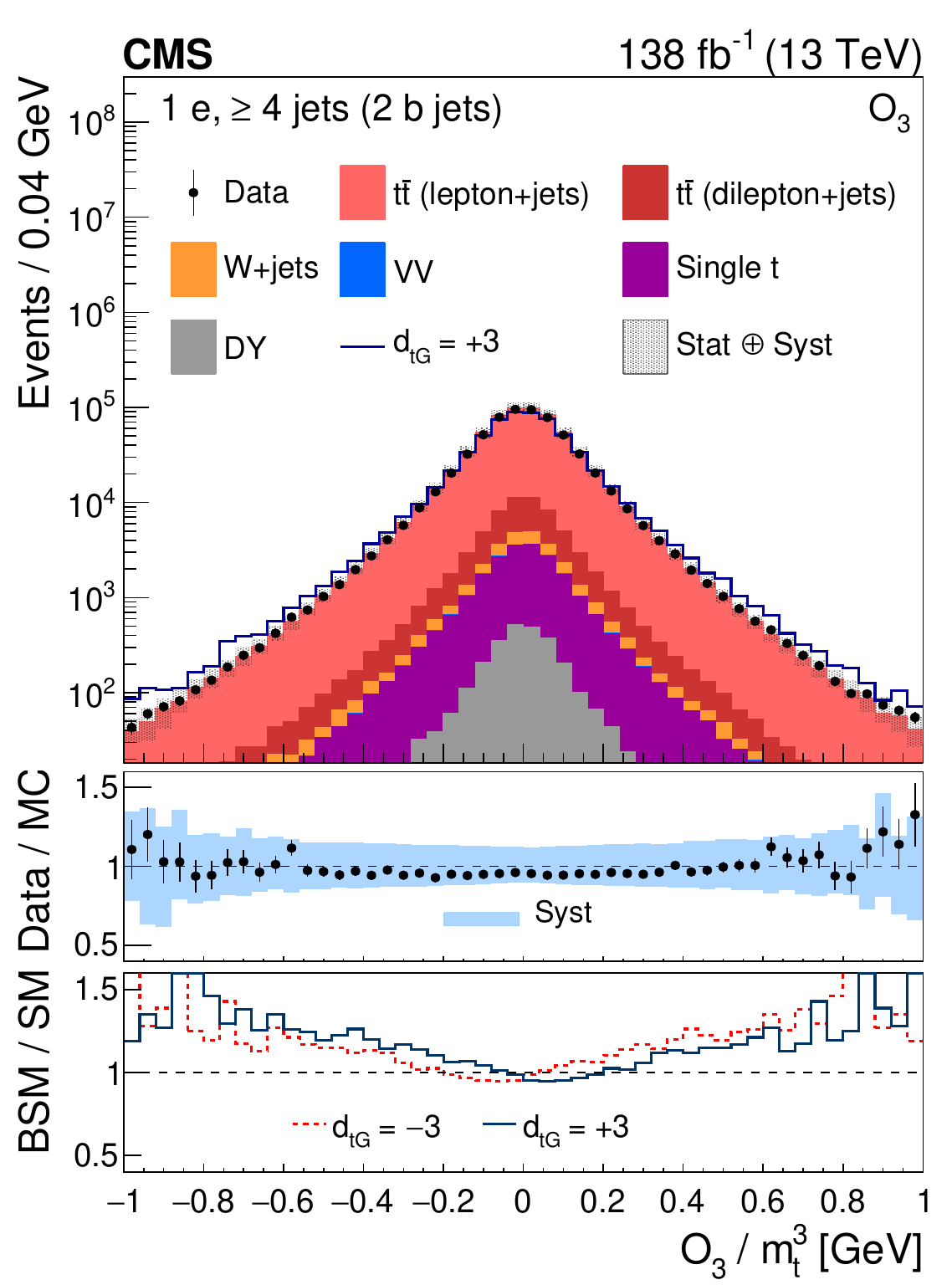}
    \includegraphics[width=0.45\textwidth]{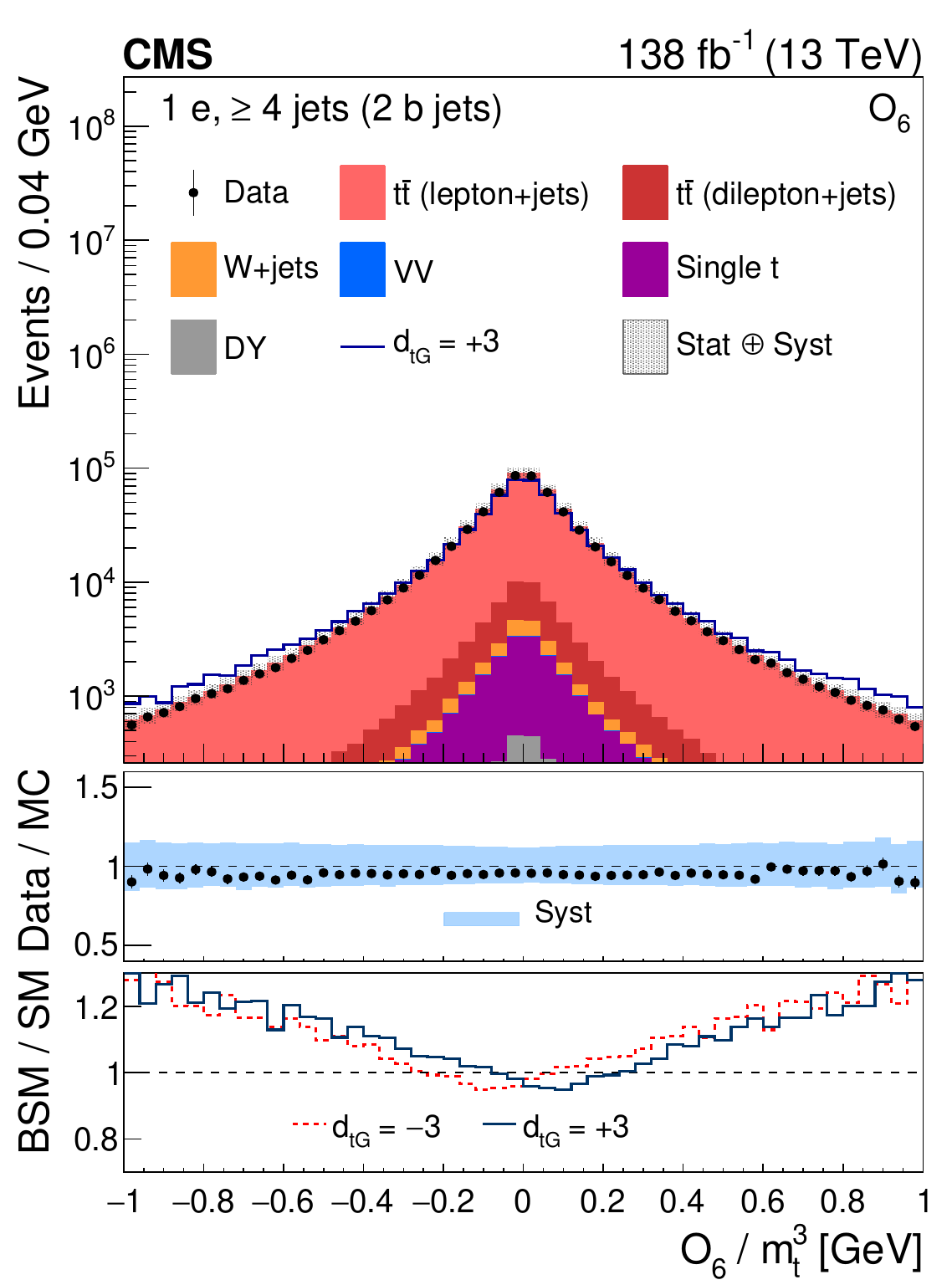}
    \includegraphics[width=0.45\textwidth]{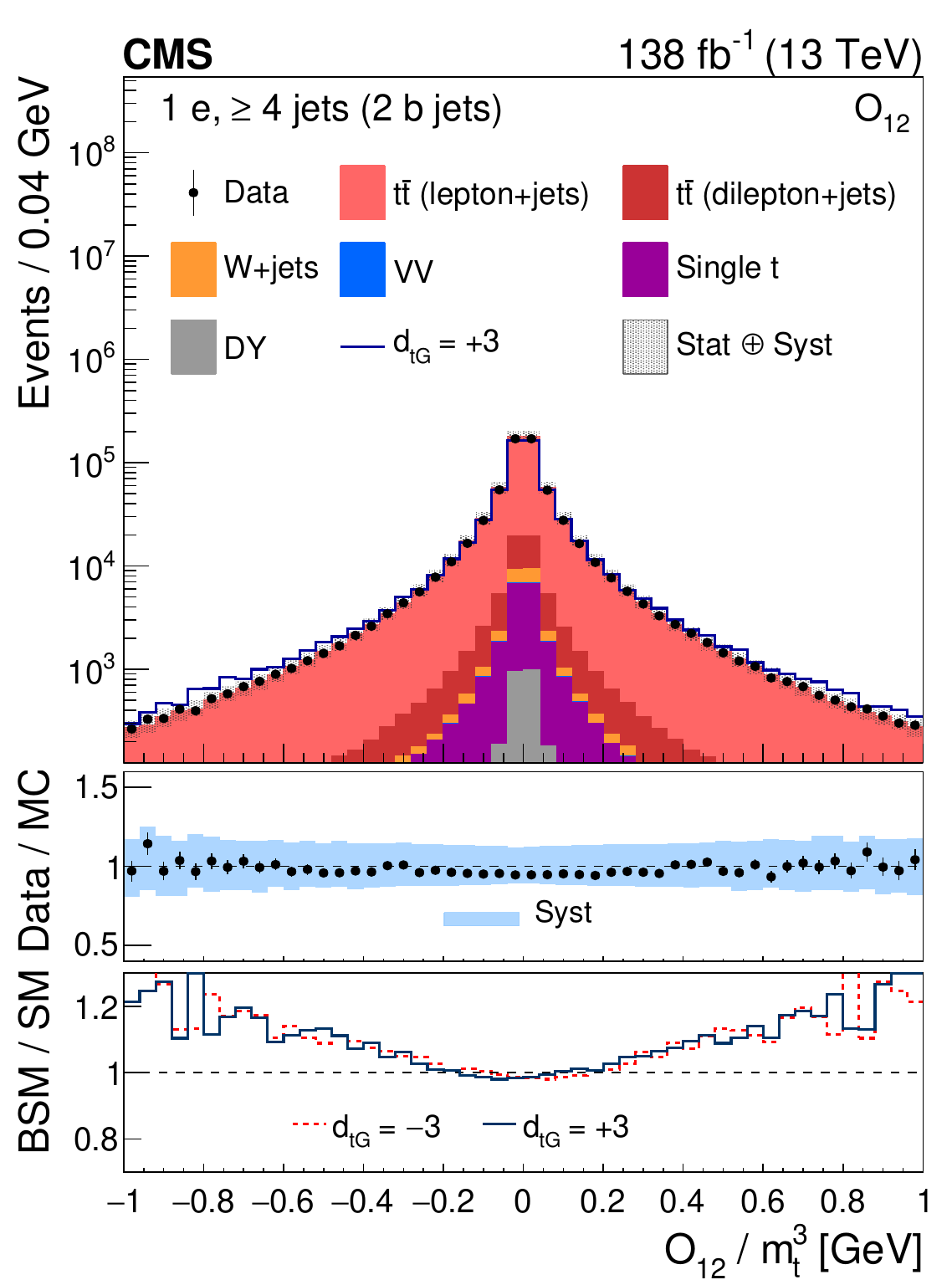}
    \includegraphics[width=0.45\textwidth]{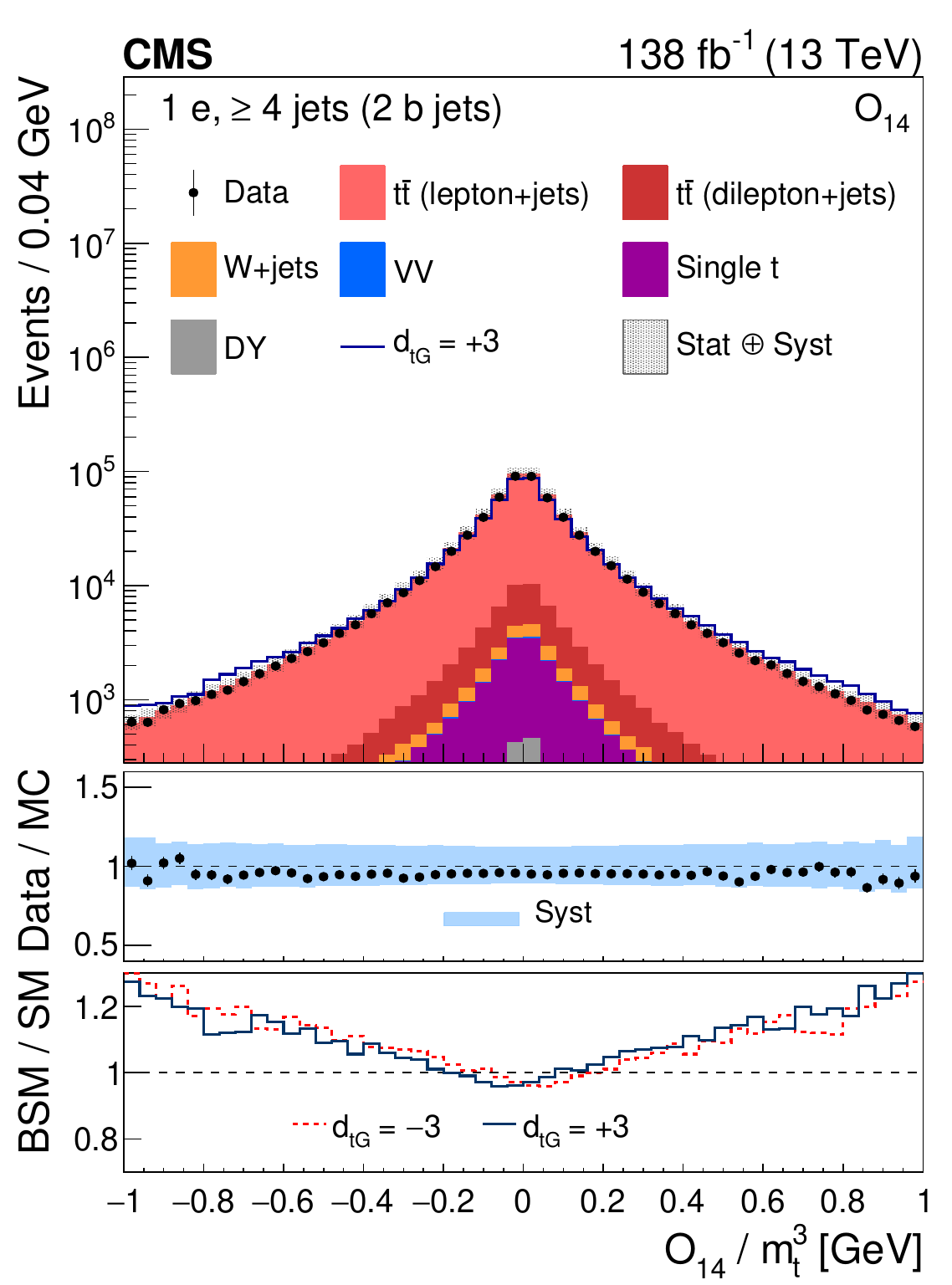}
    \caption{
        Distributions of the CP observables \Othree (upper left), \Osix (upper right), \Otwelve (lower left), and \Ofourteen (lower right), normalized with respect to \Mtcub, from data (points) and from the various sources in simulation (colored histograms) for electron events in the signal region.
        The solid-blue line shows the CEDM simulated signal normalized to the data with the CP-odd parameter $\dtG = +3$.
        The vertical bars on the data points indicate the statistical uncertainties in the data, and the hatched bands show the quadrature sum of the statistical and systematic uncertainties in the simulation.
        The lower two panels display the ratio of the data to the sum of the MC predictions and the ratio of the CEDM to the SM predictions for $\dtG = +3$ (dark-blue lines) and $-3$ (red lines).
    }
    \label{fig:el_obs_dist}
\end{figure}

\begin{figure}[!p]
    \centering
    \includegraphics[width=0.45\textwidth]{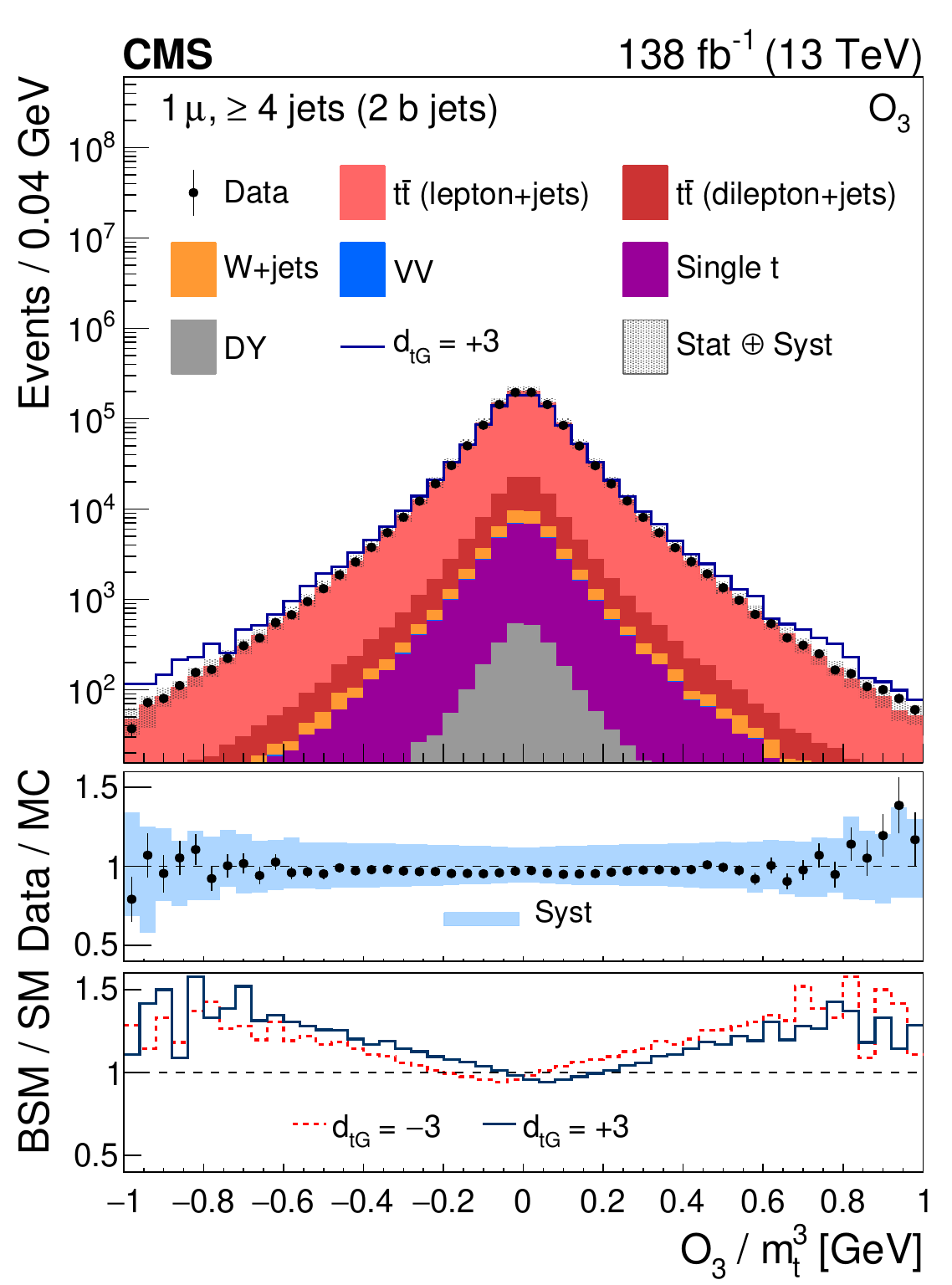}
    \includegraphics[width=0.45\textwidth]{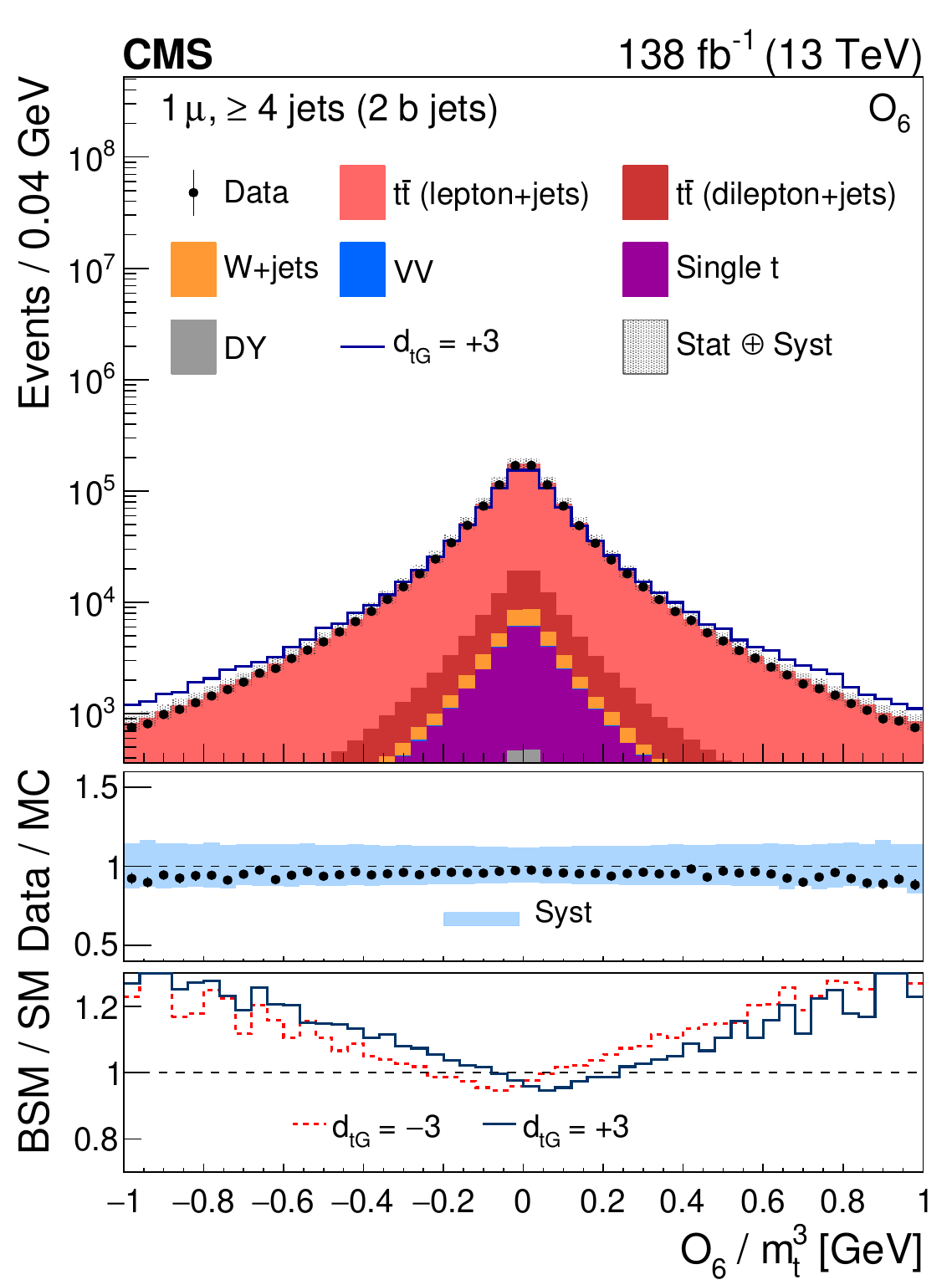}
    \includegraphics[width=0.45\textwidth]{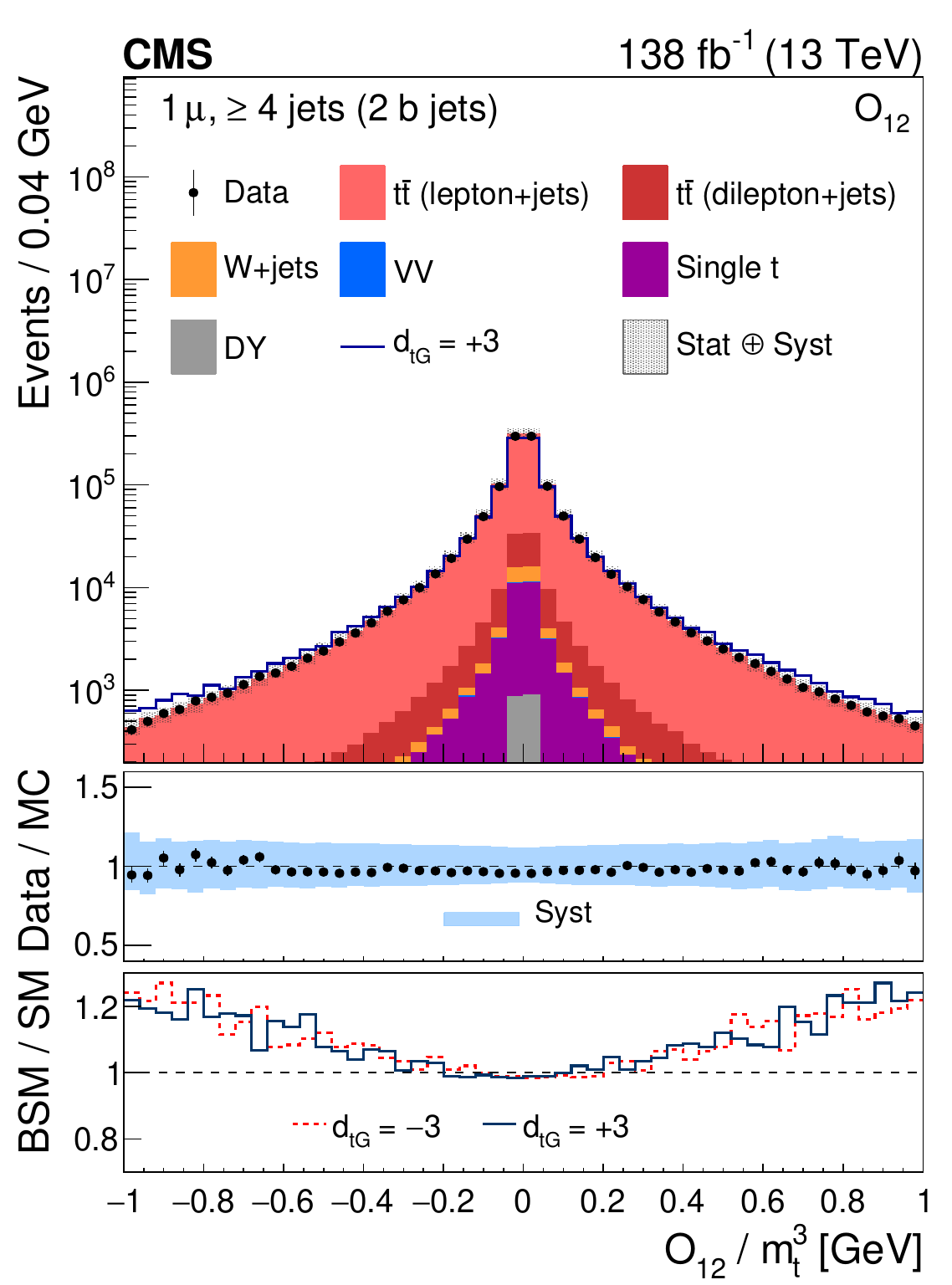}
    \includegraphics[width=0.45\textwidth]{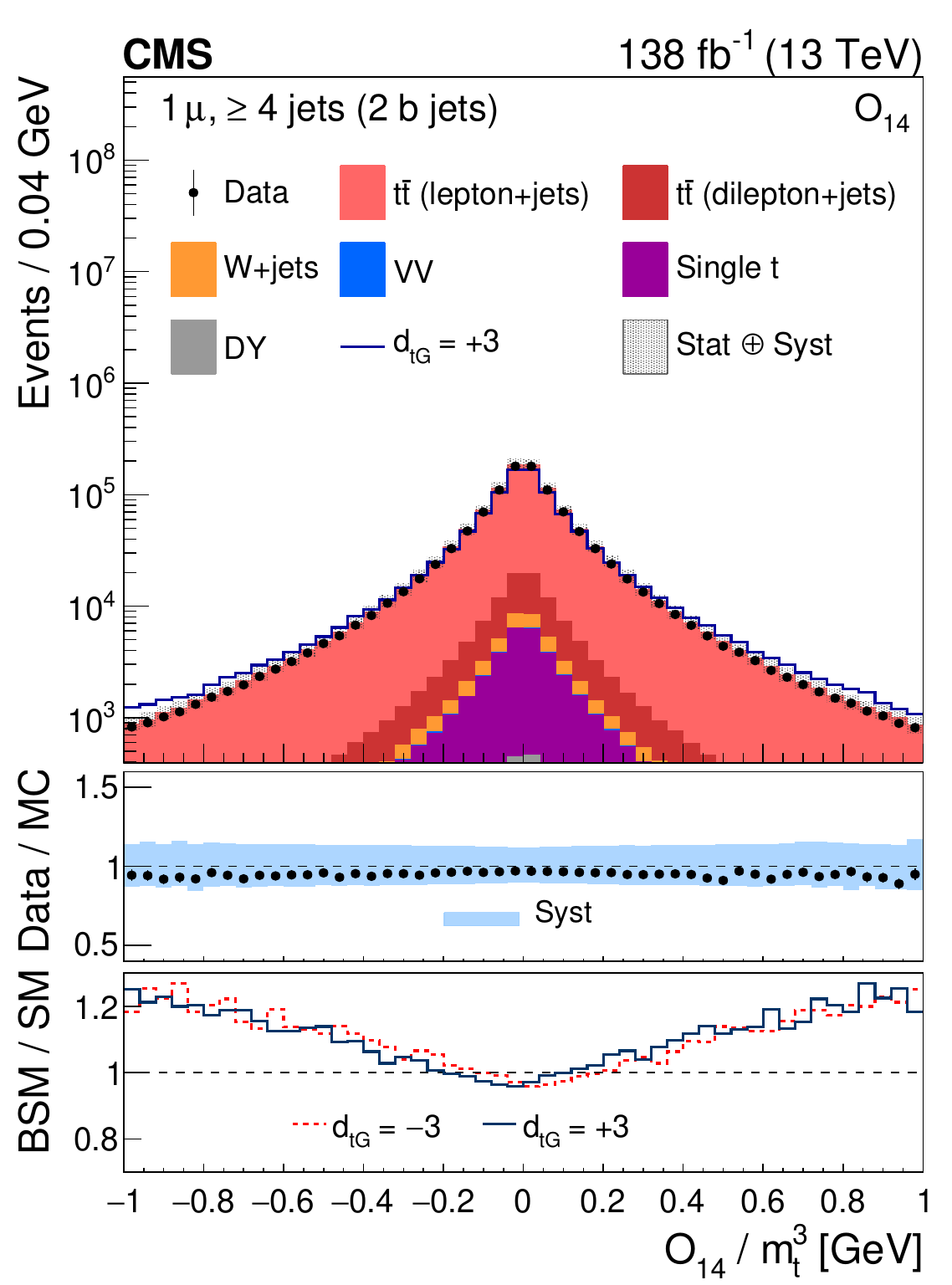}
    \caption{
        Distributions of the CP observables \Othree (upper left), \Osix (upper right), \Otwelve (lower left), and \Ofourteen (lower right), normalized with respect to \Mtcub, from data (points) and from the various sources in simulation (colored histograms) for muon events in the signal region.
        The solid-blue line shows the CEDM simulated signal normalized to the data with the CP-odd parameter $\dtG = +3$.
        The vertical bars on the data points indicate the statistical uncertainties in the data, and the hatched bands show the quadrature sum of the statistical and systematic uncertainties in the simulation.
        The lower two panels display the ratio of the data to the sum of the MC predictions and the ratio of the CEDM to the SM predictions for $\dtG = +3$ (dark-blue lines) and $-3$ (red lines).
    }
    \label{fig:mu_obs_dist}
\end{figure}

The other \PQb-tagged jet associated with the leptonically decaying \PW boson candidate is assigned as being from the hadronization of a bottom quark or antiquark based on the charge sign of the lepton.
The \PQb jets are correctly assigned in 60\% of \ttbar events.
However, from simulation it is found that when the invariant mass of the combined isolated lepton and associated \PQb-tagged jet (\Mlb) is greater than 150\GeV, the events have a large fraction of incorrect \PQb jet assignments.
Therefore, from studies of the simulated \ttbar sample, further requirements of $\chisq < 20$ and $\Mlb < 150\GeV$ are imposed.
This improves the fraction of correctly assigned \PQb jets to $\approx$74\%, while keeping $\approx$65\% of the \ttbar events.
After these selection criteria, the purity of the \ttbar events is 95\%, with single top quark production contributing a background of 3\%.
The upper panels in Figs.~\ref{fig:el_obs_dist} and~\ref{fig:mu_obs_dist} show the distributions of the CP observables for the electron and muon channels, respectively.
The middle panels display the ratio of the data to the simulated distributions, which show reasonable agreement within the one-standard-deviation band.
The bottom panels present the ratio of the CEDM to the SM predictions for $\dtG = \pm 3$.
In these figures, the values of the CP observables are divided by \Mtcub to convert the units to \GeVns.
The systematic uncertainties shown in Figs.~\ref{fig:el_obs_dist} and~\ref{fig:mu_obs_dist}, as well as in later figures and tables, include all systematic uncertainties, except for those due to backgrounds.
The latter uncertainty contributes only to the final \Acp measurements instead of the number of signal events and will be discussed in detail in Section~\ref{sec:uncertainty}.
Table~\ref{tab:signal_region_expected_percentage} shows the predicted signal and background contributions to the signal events from simulation for the electron and muon channels.
The fraction of leptonic tau decays in the SM simulated \ttbar samples is about 0.3\% and the contribution is negligible in the final results.
The QCD and \ttbar multijet events are highly suppressed by the selection requirements and provide negligible contributions of around 0.2 and 0.05\%, respectively, to the signal events.
The estimation of the number of misidentified signal events coming from \ttbar decays to dilepton+jets is discussed in Section~\ref{sec:fitresult}.

\begin{table}[h!]
    \topcaption{
        The predicted \ttbar signal and background contributions to the signal events from simulation for the electron and muon channels.
    }
    \label{tab:signal_region_expected_percentage}
    \centering\renewcommand\arraystretch{1.2}
    \begin{tabular}{ccc}
        Process & Electron channel (\%) & Muon channel (\%) \\
        \hline
        \ttbar in lepton+jets & 89.9 & 89.5\\
        \ttbar in dilepton+jets & 5.5 & 5.5\\
        \ttbar multijet & 0.1 & 0.1\\
        Single \PQt & 2.9 & 2.9\\
        \Wjets & 1.0 & 1.1\\
        DY+jets & 0.4 & 0.2\\
        QCD multijet & 0.2 & 0.6\\
        $\PZ\PZ$ / $\PW\PW$ / $\PW\PZ$ & 0.1 & 0.1\\
    \end{tabular}
\end{table}

Background events can induce spurious measurements of the asymmetries, so a sample of background-enriched data events is used to check for such effects.
In order to enhance the fraction of background events and minimize the contribution from \ttbar, events are required to have no \PQb-tagged jets.
The \PQb jet veto is defined using a weakly restrictive working point of \DeepCSV, corresponding to an 84\% efficiency in identifying \PQb jets and an 11\% misidentification rate for light-flavor quark and gluon jets~\cite{CMS:bjet_eff}.
The isolation requirement in the looser set of selection criteria for leptons is relaxed, so events with additional leptons produced through the decay of a bottom quark can be more highly rejected.
The rest of the selection criteria are the same as used for the signal events.
The object assigned by the $\chisq$ algorithm takes the role of the \PQb jet coming from the hadronically decaying top quark, and the jet closest to the isolated lepton is assigned as the other \PQb jet.
The background-enriched events are expected to be dominated by non-\ttbar processes ($\approx$90\%), including a major contribution from \Wjets.

\section{Fitting procedure and extraction of asymmetries}\label{sec:fitresult}
\subsection{Fitting procedure}
The presence of CPV can manifest itself through a nonzero value of the asymmetry defined as
\begin{linenomath}\begin{equation}\label{eq:counting_method}
    \Acp(\Oi) = \frac{\Npos-\Nneg}{\Npos+\Nneg},\quad i=3,\,6,\,12,\,14.
\end{equation}\end{linenomath}
The CP-violating asymmetries $\Acp(\Oi)$ are expected to vanish in the SM.
However, nonzero CP-violating couplings of the top quark from BSM phenomena can lead to sizable asymmetries.
An anomalous CEDM contribution~\cite{CPVtop:13TeVRef} can be as large as 8 and 0.4\% for $\Acp(\Othree)$ and $\Acp(\Otwelve)$, respectively.

Experimental factors, such as the misreconstruction of the physical objects, can affect the measurements of the asymmetries~\cite{CPVtop:13TeVRef}.
For example, misidentified signal events coming from \ttbar in the dilepton+jets channel can cause spurious asymmetry measurements.
We denote \Acp as the asymmetry that would be measured with an ideal detector and \Acpprime as the measured effective asymmetry, including experimental factors.
An estimate of \Acp can be obtained after correcting the measured asymmetry for instrumental effects.
Given that the SM predicts negligible \Acp in the top quark sector, a nonzero effective \Acpprime would be a strong hint of BSM phenomena.
For this reason, measurements of \Acpprime can be computed using \ttbar events and are the primary results presented in this paper.

Because \ttbar multijet events are highly suppressed by the signal-event requirements, only the \ttbar to lepton+jets and dilepton+jets channels are considered in the \ttbar contribution, with the latter assumed to be background.
The signal and background yields are determined through an extended maximum likelihood fit to the \Mlb distributions in data using simulated-event templates.
The \ttbar template is obtained from simulation, and the background template from the background-enriched events in data.
For each CP observable, the templates are classified according to the sign of the CP observable.
The extended likelihood function is defined as
\begin{linenomath}\begin{equation}\begin{aligned}
    L_{\text{ext}} &= \frac{\re^{-(\numpos+\ratiob \numb)}}{\Numpos!}\prod^{\Numpos}_{i=1} \numpos \fspos(\Mlbi) + \ratiob \numb \fbpos(\Mlbi) \\
    &+ \frac{\re^{-(\numneg+(1-\ratiob) \numb)}}{\Numneg!}\prod^{\Numneg}_{i=1} \numneg \fsneg(\Mlbi) + (1-\ratiob) \numb \fbneg(\Mlbi),
\end{aligned}\end{equation}\end{linenomath}
where \numb, \numpos, and \numneg are the parameters of interest and refer to the yields of background, and \ttbar events with positive and negative CP observable values, respectively; \ratiob refers to the fraction of background events with positive CP observable values obtained from the background template; $\fspos(\Mlb)$, $\fsneg(\Mlb)$, $\fbpos(\Mlb)$, and $\fbneg(\Mlb)$ refer to the probability density functions (pdf) of \Mlb obtained from the \ttbar and background templates according to the sign of the CP observable, respectively; and \Numpos and \Numneg are the number of events according to the sign of the CP observable.
To improve the fit results, the upper bound on \Mlb is relaxed to 500\GeV in the fit in order to include more sideband events.
However, to be consistent with the signal-event requirements, events with $\Mlb > 150\GeV$ are excluded after the fit in determining the final event yields for each CP observable.

\begin{figure}[!t]
    \centering
    \includegraphics[width=0.45\textwidth]{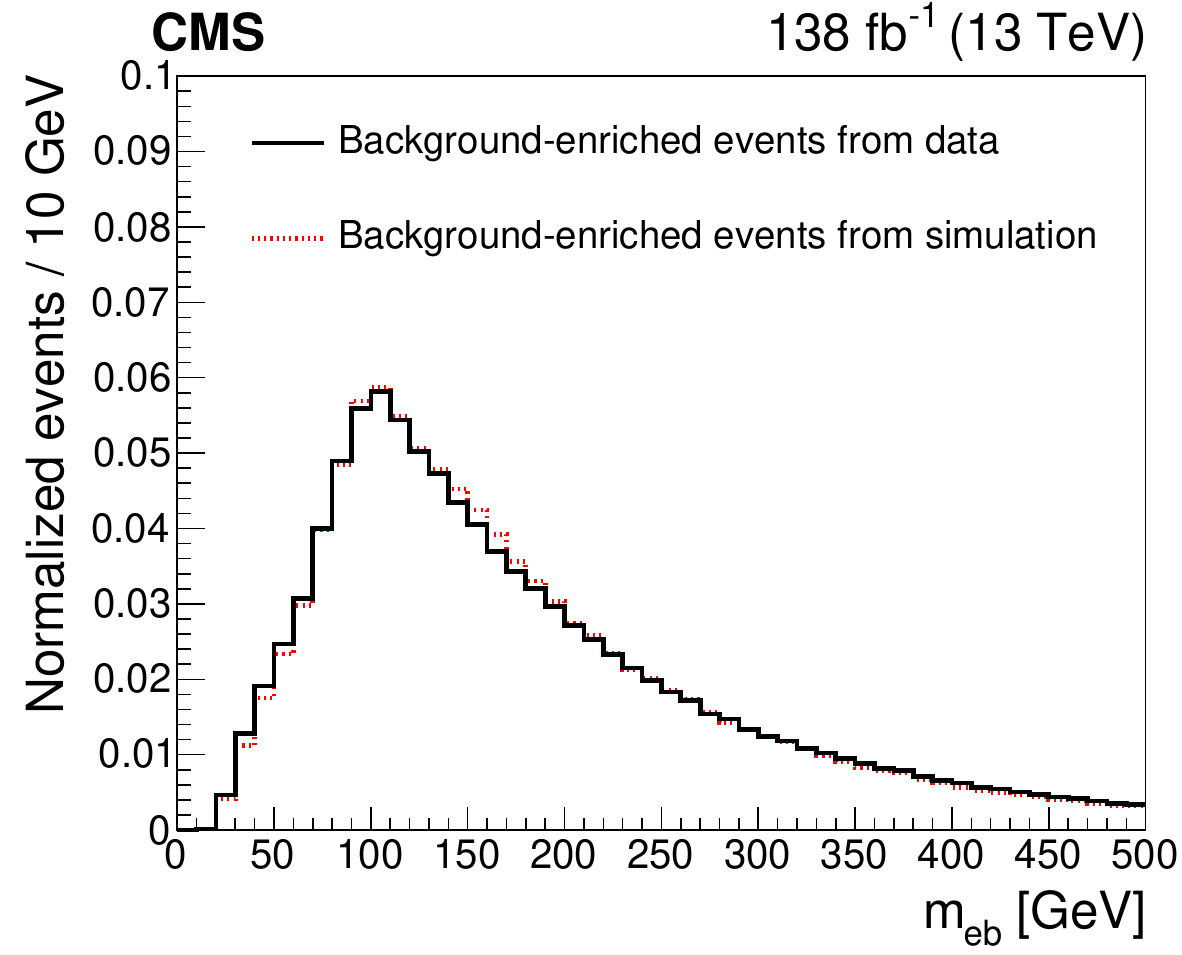}
    \includegraphics[width=0.45\textwidth]{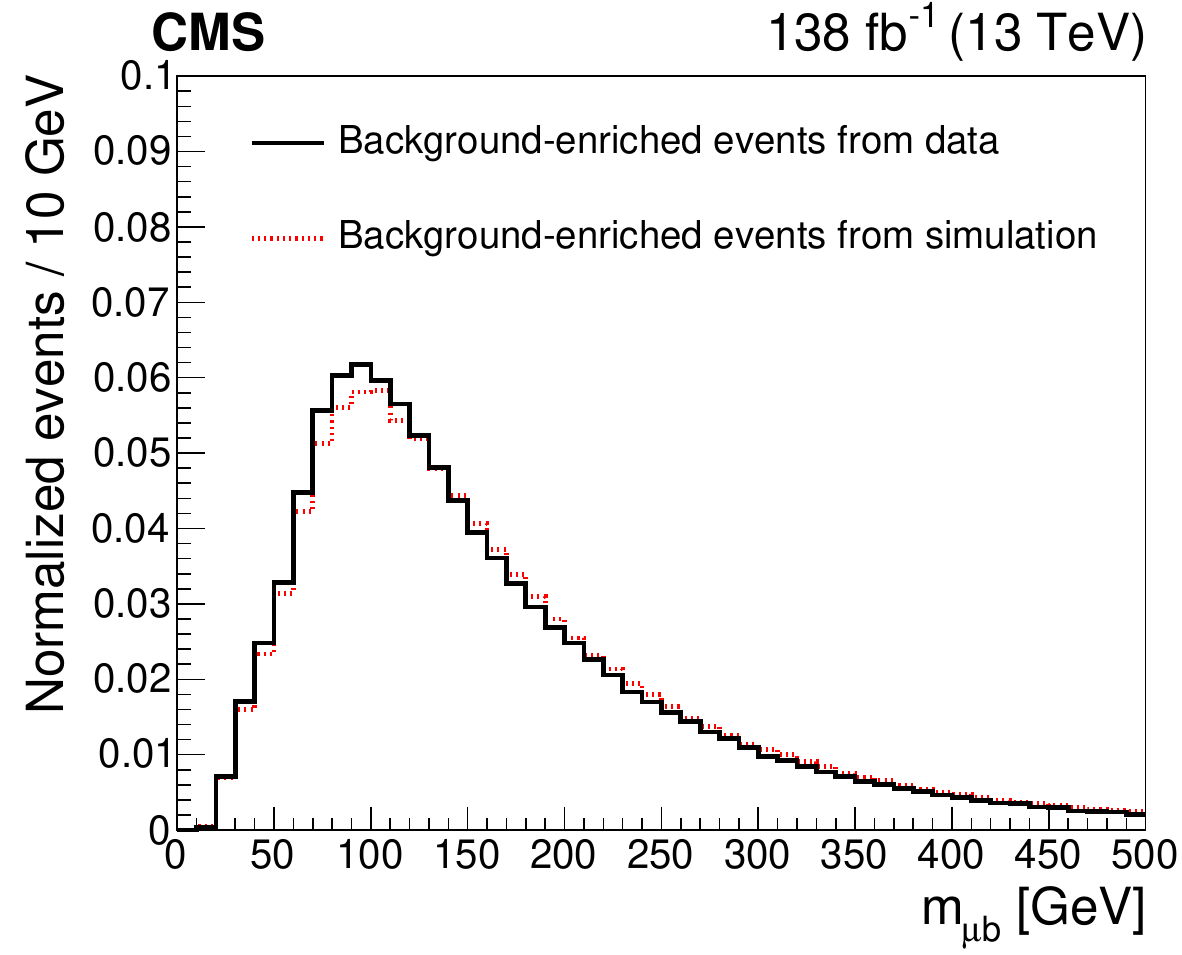}
    \includegraphics[width=0.45\textwidth]{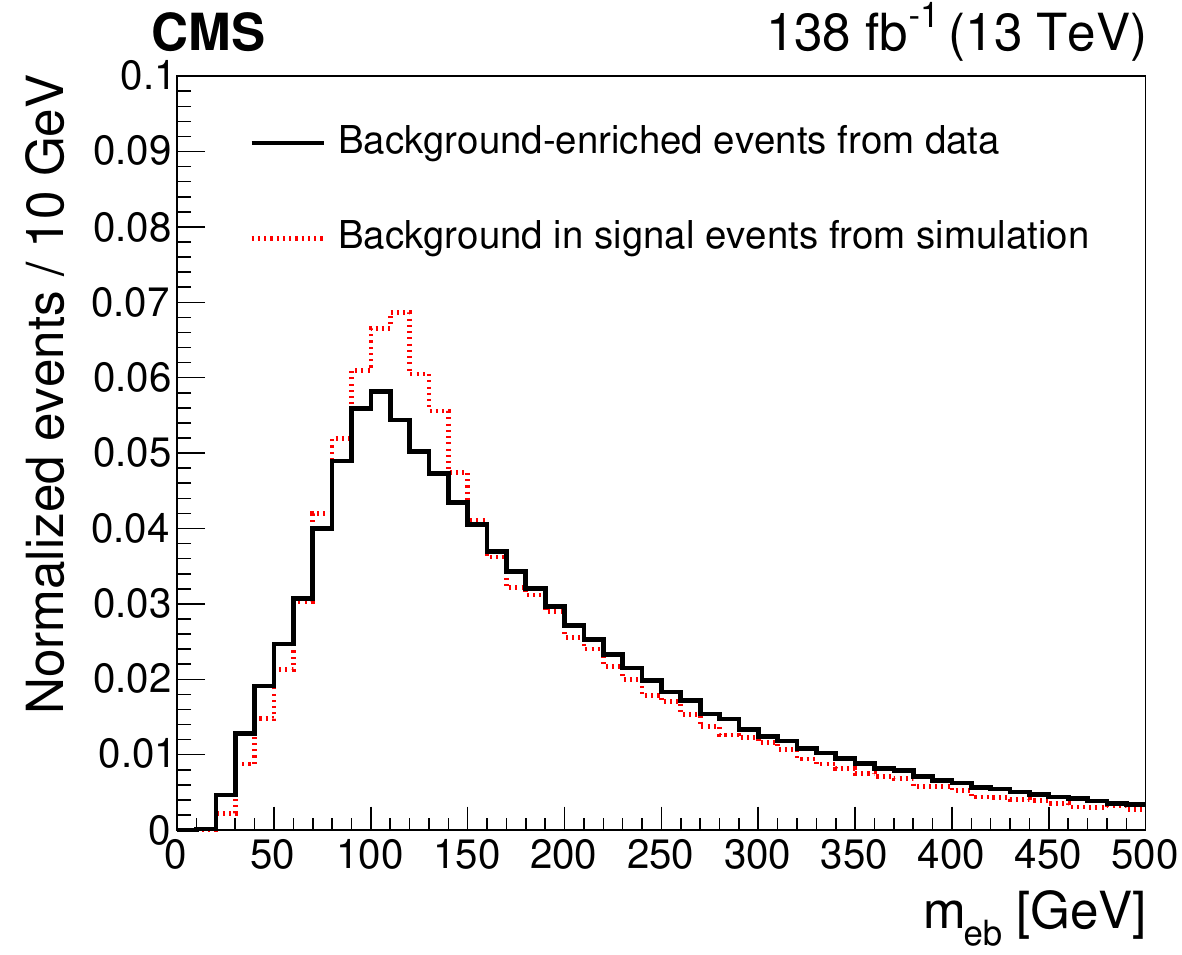}
    \includegraphics[width=0.45\textwidth]{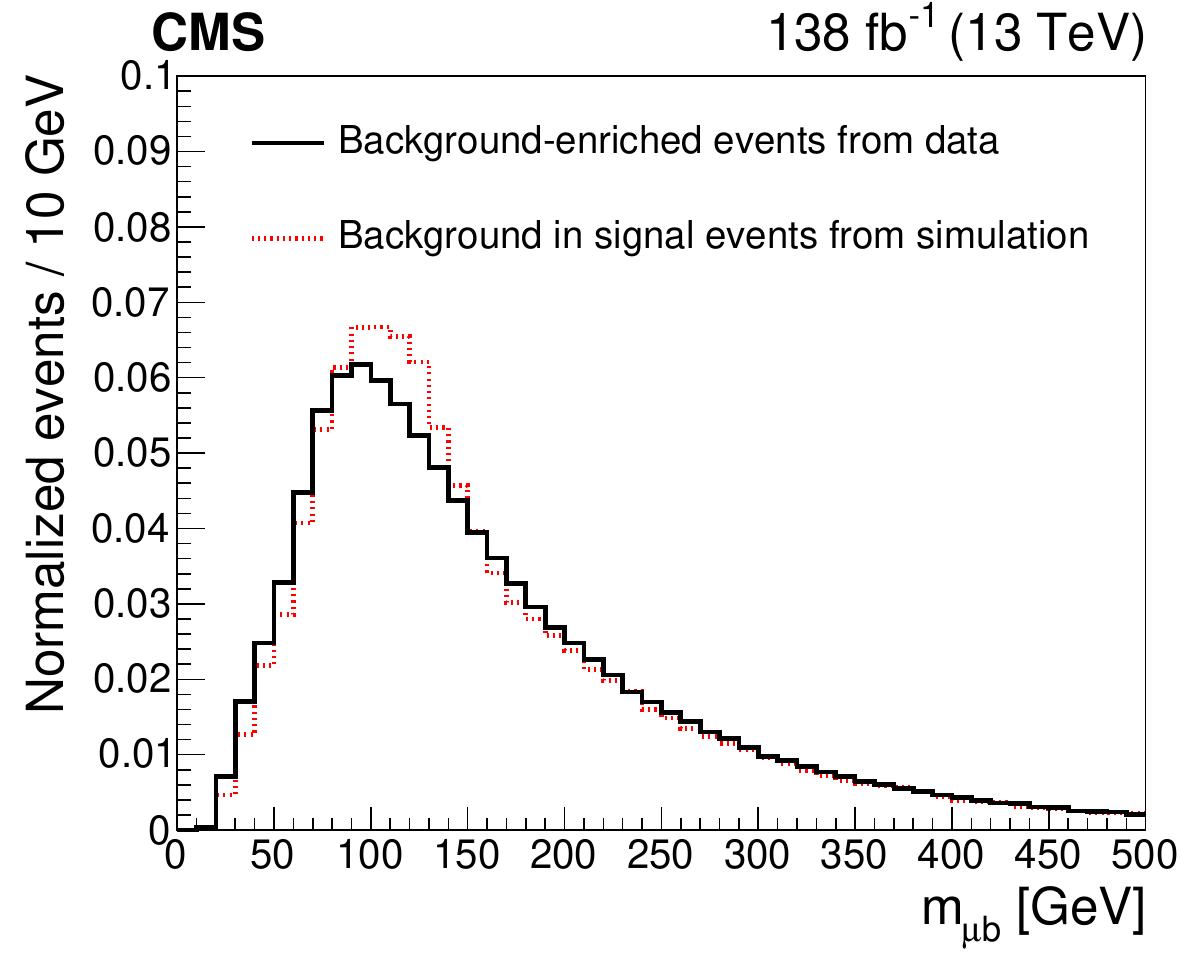}
    \caption
    {
        The normalized \Mlb distributions for the electron (left) and muon (right) channels.
        The upper two plots compare the background-enriched distributions from data (solid line) to the MC predictions (dotted-red line).
        The lower two plots give the background-enriched distributions from data (solid line) and the MC predictions for the distributions from the background in the signal events.
    }
    \label{fig:CR_SR_closure_test}
\end{figure}

Figure~\ref{fig:CR_SR_closure_test} displays the normalized \Mlb distributions from the signal and background-enriched events for the electron (left) and muon (right) channels.
The upper plots compare the distribution for the background-enriched data events to that from the MC prediction.
Good agreement is observed between the two distributions, showing that the background-enriched events are consistent with being entirely background.
The lower plots display the distributions for the background-enriched events in data and the predicted MC background in the signal-event sample.
The \Mlb distribution of background-enriched events in data is slightly wider than the predicted MC background in signal events.
This difference is taken as one of the systematic uncertainties, as discussed in Section~\ref{sec:uncertainty}.

\begin{figure}[!t]
    \centering
    \includegraphics[width=0.45\textwidth]{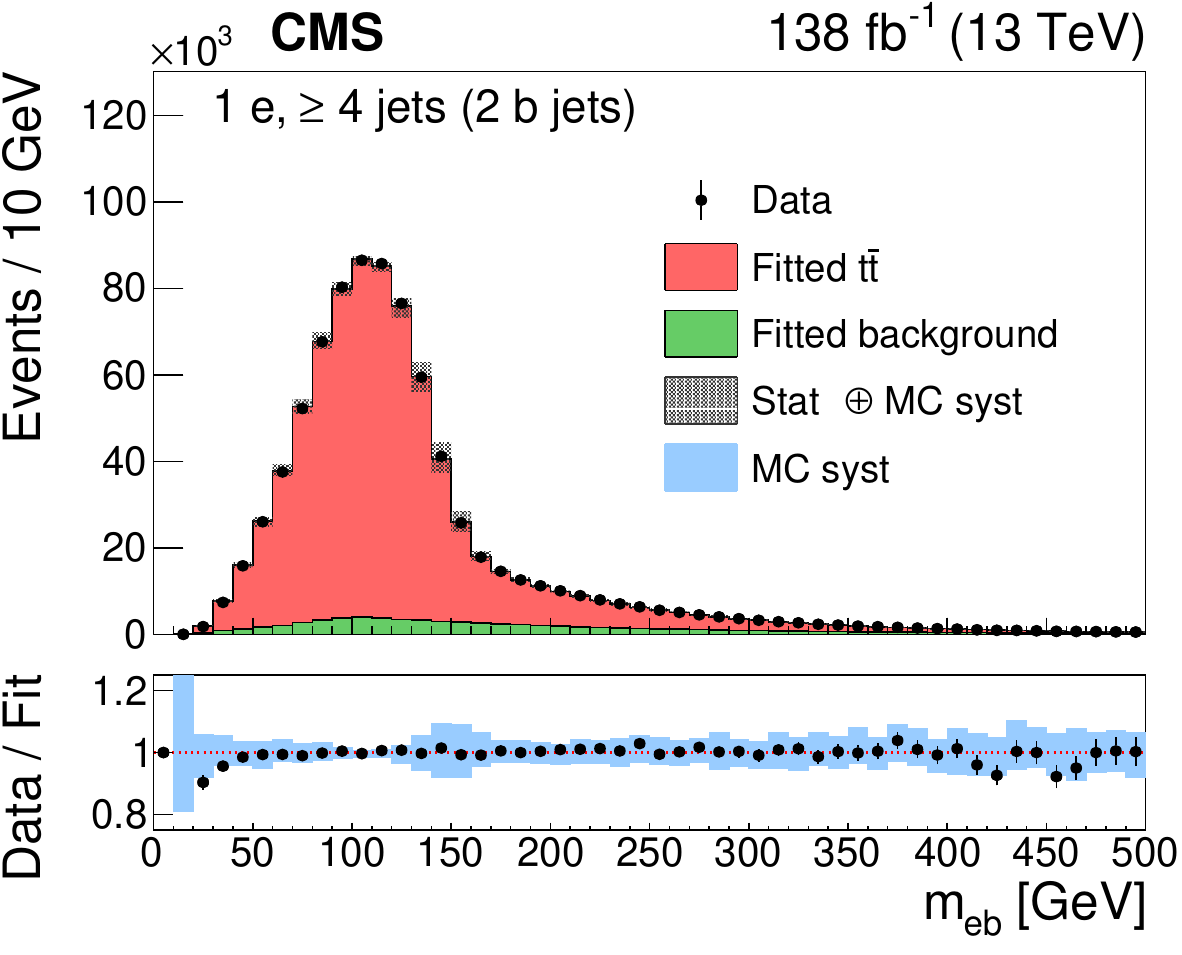}
    \includegraphics[width=0.45\textwidth]{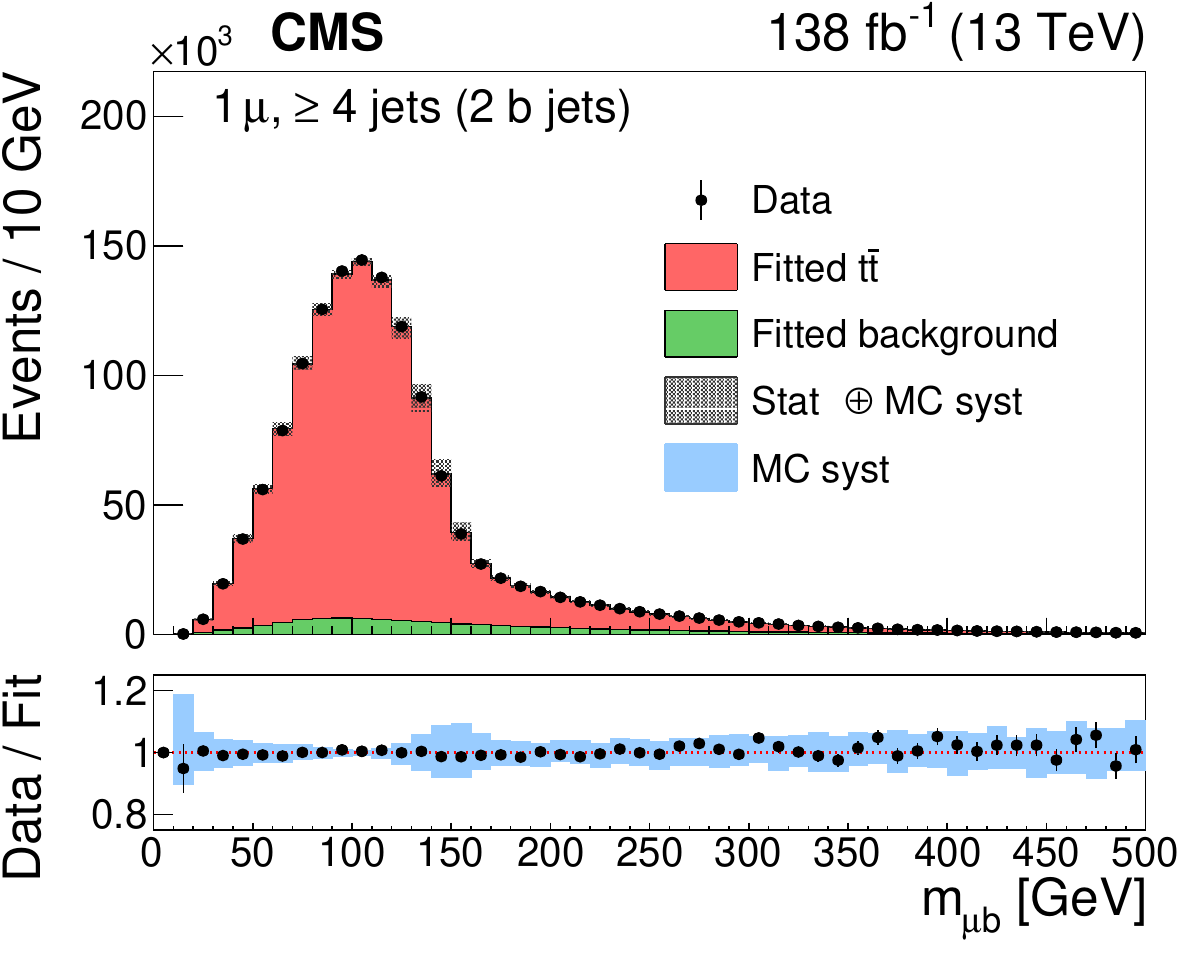}
    \caption{
        The \Mlb invariant mass distributions in the electron (left) and muon (right) channels from data (points).
        The results of the fit to the \ttbar and background templates are shown by the red and green histograms, respectively.
        The vertical bars on the data points in the upper panels indicate the statistical uncertainties in the data and the hatched bands show the combined statistical and systematic uncertainties in the simulation.
        The lower panels give the ratio of the data to the sum of the fitted MC predictions.
        The blue bands represent the systematic uncertainties in the expected yield in the simulation for all sources of systematic uncertainty (Section~\ref{sec:uncertainty}).
    }
    \label{fig:fitting_results}
\end{figure}

The \Mlb distributions in data are shown in Fig.~\ref{fig:fitting_results}, along with the results of the fit.
The measured numbers of \ttbar signal and background events in the electron and muon channels from the fits, and the corresponding \ttbar purities, are presented in Table~\ref{tab:fitting_purity}.
The final \Acpprime measurements can then be computed using the event-counting method of Eq.~\eqref{eq:counting_method}.

\begin{table}[!h]
    \topcaption
    {
        The fitted number of \ttbar signal and \ttbar background events (fitted \ttbar) and other background events (fitted background) in the electron and muon channels, along with the \ttbar purities.
        Although the fit is performed for $\Mlb < 500\GeV$, the event yields are given for $\Mlb < 150\GeV$.
        The uncertainties shown are statistical only.
    }
    \label{tab:fitting_purity}
    \centering\renewcommand\arraystretch{1.2}
    \begin{tabular}{ccc}
        & Electron channel & Muon channel\\
        \hline
        Fitted \ttbar & $604\,700 \pm 1200$ & $1\,062\,600 \pm 1500$\\
        Fitted background & $34\,030 \pm 480$ & $58\,490 \pm 820$\\
        Fitted fraction of \ttbar (\%) & $94.7 \pm 0.1$ & $94.8 \pm 0.1$ \\
    \end{tabular}
\end{table}

\subsection{Experimental sensitivity}
The sensitivity of the asymmetry measurement can be diluted due to detector and reconstruction effects, which can be parametrized through a dilution factor (\dilution).
The \Acpprime and \Acp values are related through \dilution, applied as a multiplicative correction $\Acpprime = \dilution \Acp$~\cite{CPVtop:13TeVRef}.
The dilution factor can be defined as
\begin{linenomath}\begin{equation}
    \dilution = \epsc - \epsw,
\end{equation}\end{linenomath}
where \epsc is the fraction of events where the measured CP observable has the correct sign, and \epsw is the fraction with the wrong sign.
Events are classified into the correct-sign (wrong-sign) type when the sign of the CP observable at the reconstruction level agrees with (differs from) that at the \POWHEG generator level.

The value of \dilution depends on the final-state objects used to form each CP observable.
In contrast to \Otwelve and \Ofourteen, it is necessary for \Othree and \Osix to distinguish the charges of the \PQb quarks by using the sign of the lepton charge, and therefore they tend to have a lower fraction of correct-sign events and smaller values of \dilution.
The value of \dilution can also be affected by possible contributions of BSM processes, because the CP observables are reconstructed using kinematic features of the final-state objects.
The values of \dilution determined from the SM simulations are given in Table~\ref{tab:dilution_factor}, along with their systematic uncertainties described in Section~\ref{sec:uncertainty}.

\begin{table}[!h]
    \topcaption
    {
        The dilution factor \dilution determined from simulation and its systematic uncertainty for each CP observable.
        The statistical uncertainty is negligible compared to the systematic uncertainty.
    }
    \label{tab:dilution_factor}
    \centering\renewcommand\arraystretch{1.4}
    \begin{tabular}{cc}
        CP observable & Dilution factor \dilution \\
        \hline
        \Othree  & $0.46\,^{+0.01}_{-0.02}$\\
        \Osix  & $0.44\,^{+0.01}_{-0.02}$\\
        \Otwelve & $0.74\,^{+0.01}_{-0.02}$\\
        \Ofourteen & $0.60\,^{+0.01}_{-0.01}$\\
    \end{tabular}
\end{table}

\begin{figure}[h!]
    \centering
    \includegraphics[width=0.45\textwidth]{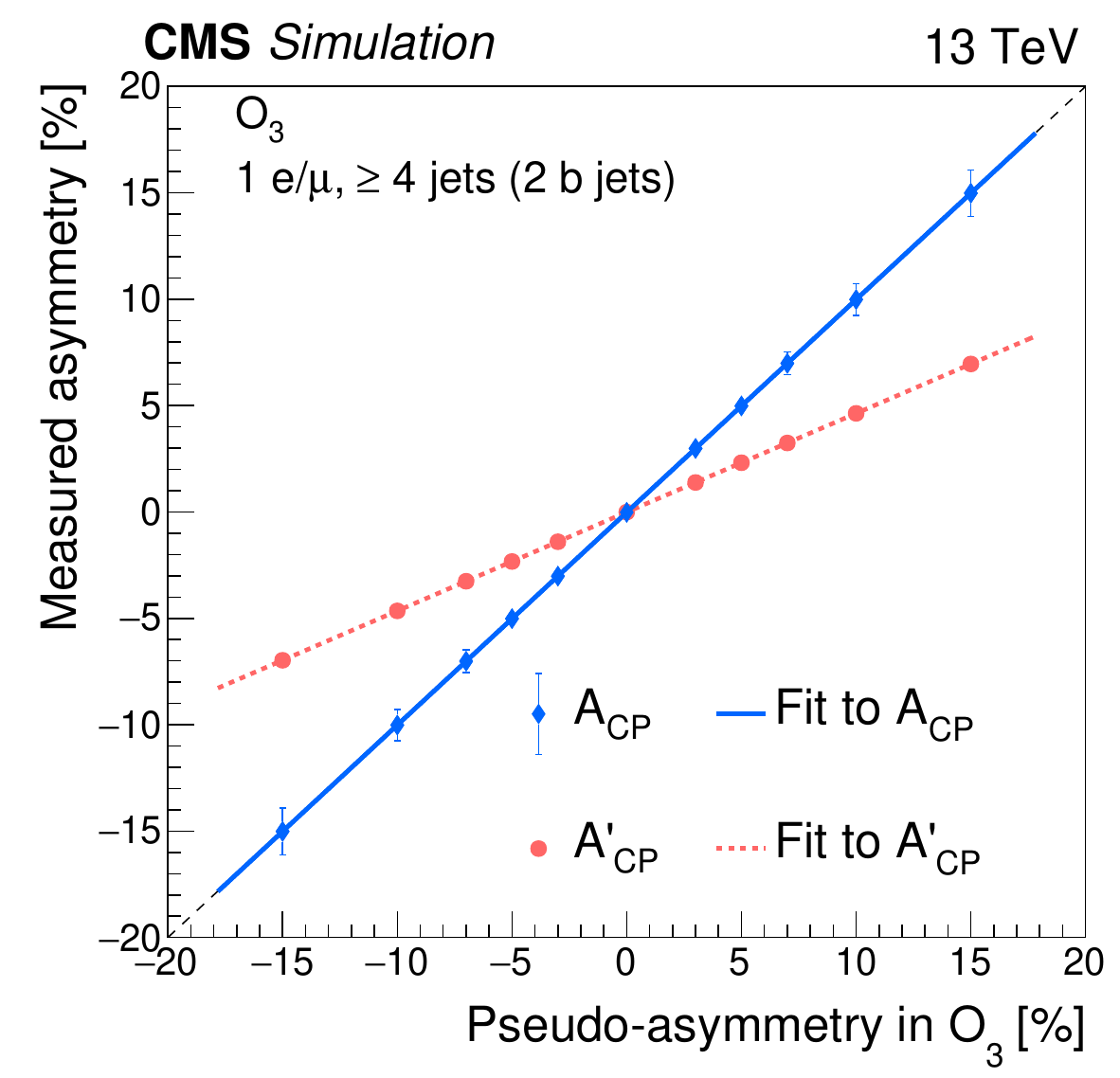}
    \includegraphics[width=0.45\textwidth]{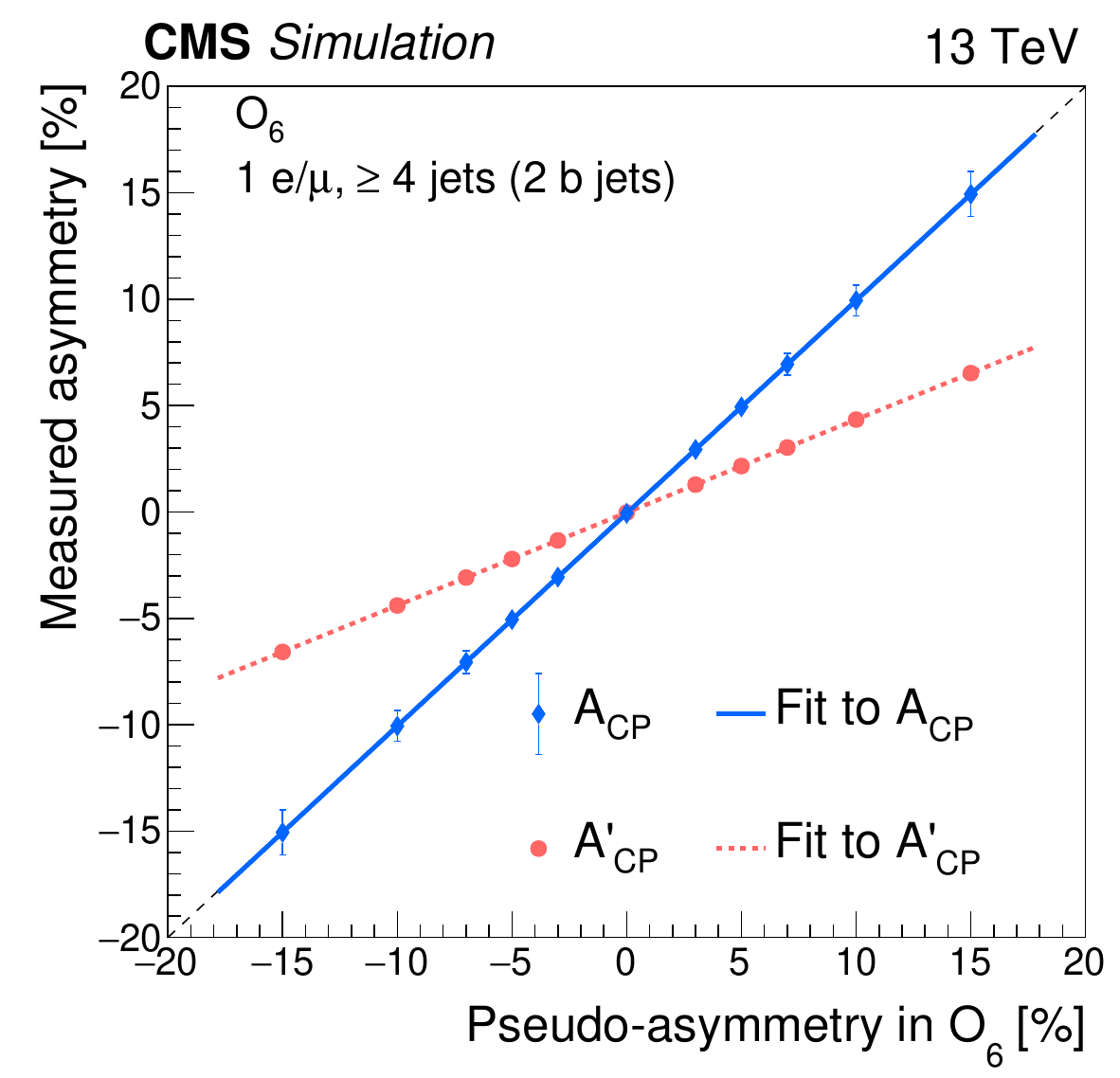}
    \includegraphics[width=0.45\textwidth]{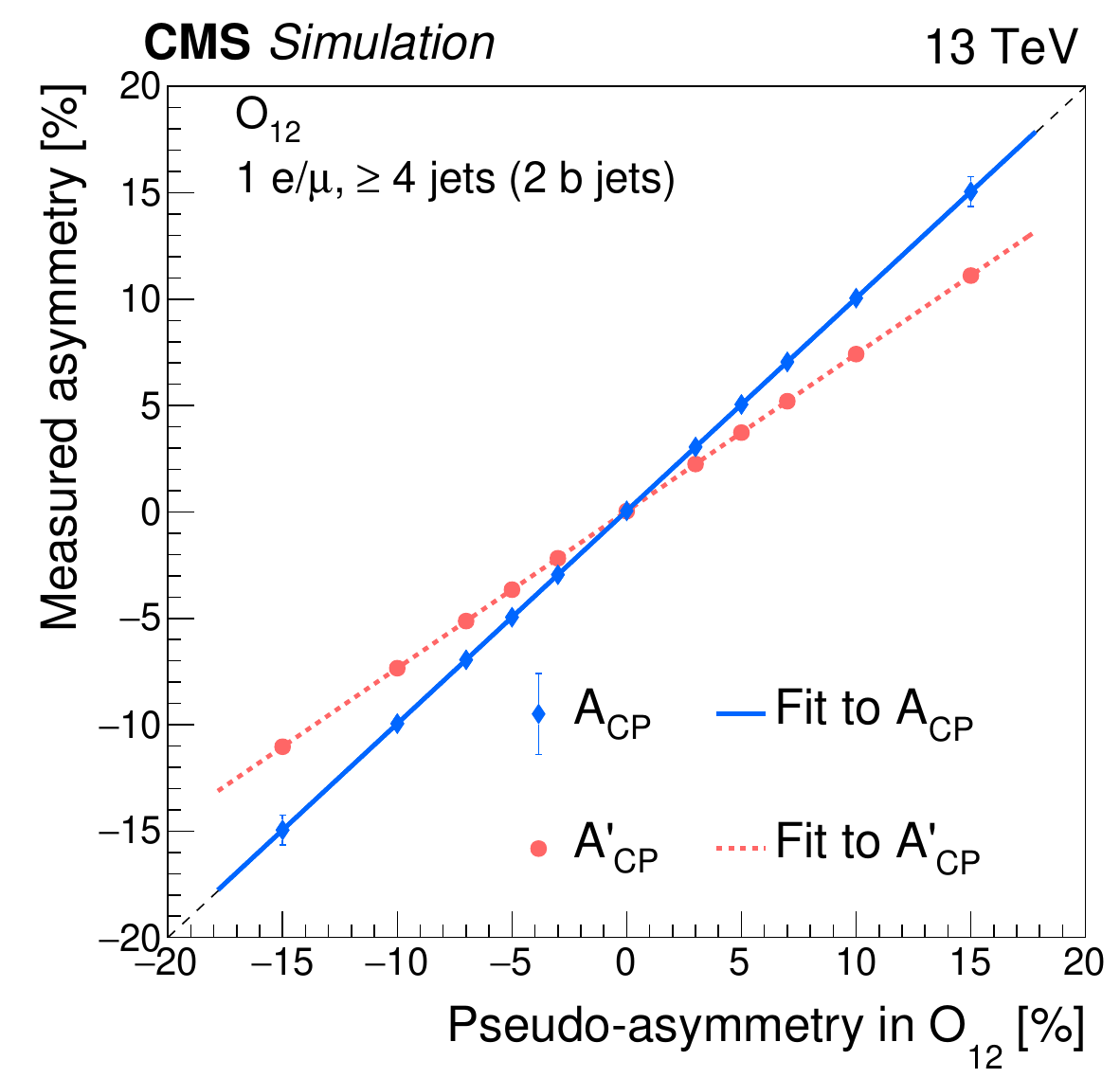}
    \includegraphics[width=0.45\textwidth]{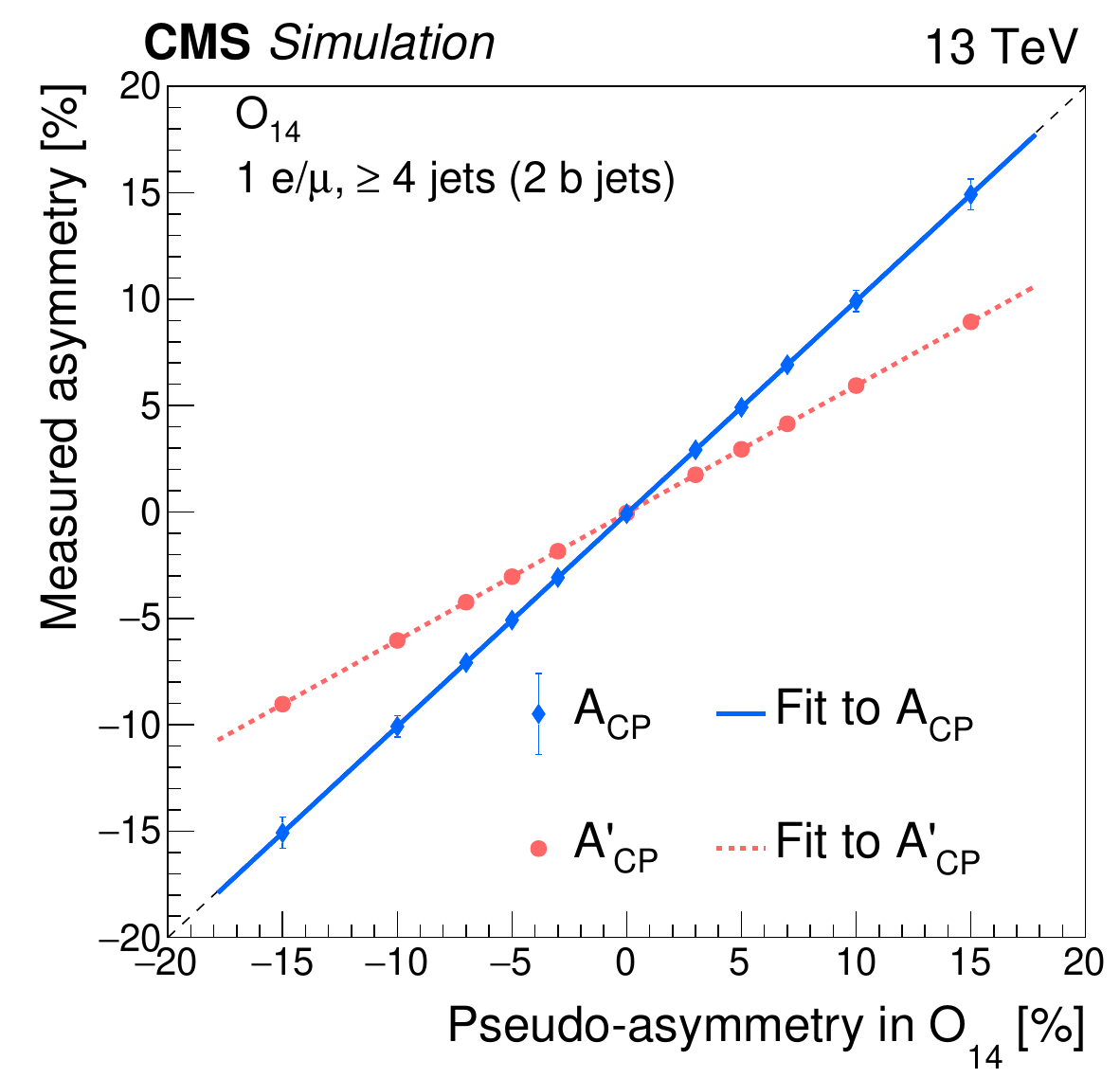}
    \caption
    {
        The measured CP-violating asymmetries in simulation as a function of the generator-level pseudo-asymmetry for the CP observables \Othree (upper left), \Osix (upper right), \Otwelve (lower left), and \Ofourteen (lower right).
        The red circles and blue diamonds give the \Acpprime and \Acp values, respectively, with the red-dotted and blue-solid lines showing the results of linear fits to those corresponding values.
        The \Acp value is obtained by dividing the \Acpprime value by the dilution factor determined from simulation for that CP observable.
        The statistical uncertainties in the \Acpprime values are smaller than the markers.
    }
    \label{fig:check_dilution_factor}
\end{figure}

In order to better understand the measured \dilution values, the lepton+jets and dilepton+jets events in the simulated \ttbar samples are reweighted at the generator level to produce various pseudo-asymmetry values.
As shown in Fig.~\ref{fig:check_dilution_factor}, the CP-violating asymmetries \Acp (diamond points) are then obtained by dividing the effective asymmetries \Acpprime (circular points) by the corresponding dilution factor determined from simulation.
Fitting the \Acpprime and \Acp values to linear functions of the generator-level pseudo-asymmetry values results in the red-dotted and blue-solid lines shown in Fig.~\ref{fig:check_dilution_factor}, respectively.
All the fits have good \chisq values, and the slopes and $y$ intercepts of the fitted lines to \Acp are consistent with 1.0 and 0.0, respectively.

\section{Systematic uncertainties}\label{sec:uncertainty}
The sources of systematic uncertainty considered in this paper are presented below in the following order: intrinsic detector bias, other experimental uncertainties, and theoretical uncertainties.
The systematic effects are expected to contribute similarly to the positive and negative regions of each CP observable, thereby largely canceling in the ratio shown in Eq.~\eqref{eq:counting_method}, used to measure the CP asymmetries.

\subsection{Detector and reconstruction effects}
The CP-violating asymmetries can arise due to both couplings from BSM processes and detector and reconstruction effects.
The previous study at \oldTeV used \Wjets samples to study the detector bias~\cite{CPVtop:CMSresult}.
The precision in the modeling of the resulting CP-violating asymmetries caused by detector and reconstruction effects may be improved by studying signal events, and therefore an event-mixing method is used to evaluate the possible bias resulting in a nonzero value of \Acpprime.
The event-mixing method is performed by mixing the four-momentum information of the \PQb-tagged jet from the hadronically decaying top quark and the highest \PT light-flavor jet among the events.
A total of 1000 mixed data sets are produced by applying the event-mixing method to the data in the signal events to eliminate possible effects from any BSM coupling.
For the $i^{\text{th}}$ mixed data set, the corresponding four-momentum information of the $j^{\text{th}}$ event is passed circularly to the $(j+i)^{\text{th}}$ event.
The resulting sets of asymmetries are independent of the physical processes involved in the signal and background events and represent only the bias from the detector and reconstruction effects.
Separate measurements are performed for each CP observable, year of data taking, and lepton flavor.
The resulting mean values of \Acpprime are presented in Table~\ref{tab:detector_bias} and show no statistically significant detector or reconstruction bias for either lepton flavor.

\begin{table}[!h]
    \topcaption
    {
        The \Acpprime values and their statistical uncertainties in percent for each CP observable from the electron and muon event-mixing samples and their combination, used to search for detector or reconstruction bias.
    }
    \label{tab:detector_bias}
    \centering\renewcommand\arraystretch{1.2}
    \begin{tabular}{cccc}
        \multicolumn{4}{c}{\Acpprime from event-mixing samples (\%)}\\
        CP observable & \ejets & \mjets & Combined\\
        \hline
        \Othree & $-0.004 \pm 0.004$ & $+0.005 \pm 0.003$ & $+0.003 \pm 0.003$\\
        \Osix & $-0.006 \pm 0.005$ & $+0.003 \pm 0.003$ & $+0.003 \pm 0.003$\\
        \Otwelve & $+0.005 \pm 0.005$ & $-0.001 \pm 0.003$ & $+0.001 \pm 0.002$\\
        \Ofourteen & $-0.008 \pm 0.004$ & $-0.001 \pm 0.003$ & $-0.005 \pm 0.002$\\
    \end{tabular}
\end{table}

\subsection{Other experimental and theoretical systematic uncertainties}
In order to avoid bias from a single estimation of a systematic uncertainty, pseudo-experiments are employed instead.
For each systematic change, a reference histogram is created using a nominal signal template.
Four thousand sets of pseudo-data are sampled from the reference histogram and fitted with the nominal (varied) signal and background templates to get the nominal (varied) fitted signal yields.
The \Acpprime can then be measured using the signal yields in the positive and negative regions of each CP observable.
The 4000 \Acpprime values for each CP observable are fit to a Gaussian function to obtain a mean and standard deviation.
The larger of the absolute value of the mean and the standard deviation is then used to estimate the systematic uncertainty from this source.
The results are summarized in Table~\ref{tab:acp_uncertainties} and described in the following subsections.

\begin{table}[!p]
    \topcaption
    {
        The sources and values of the systematic uncertainties in \Acpprime for each of the CP observables in percent, averaged over the two lepton-flavor channels.
        The experimental sources are listed first and then the theoretical ones.
    }
    \label{tab:acp_uncertainties}
    \centering\renewcommand{\arraystretch}{1.5}
    \newcommand{\TwoRow}[2]{\multirow{2}{*}{\shortstack{#1\\ #2}}}
    \newcommand{\OneRow}[1]{\multirow{2}{*}{#1}}
    \begin{tabular}{ccccc}
        Systematic sources & \multicolumn{4}{c}{\Acpprime (\%)} \\[-3pt]
        & \Othree & \Osix & \Otwelve & \Ofourteen\\
        \hline
        \OneRow{Pileup} & \TwoRow{$-0.0008$}{$+0.0010$} & \TwoRow{$-0.0003$}{$+0.0007$} & \TwoRow{$+0.0023$}{$-0.0017$} & \TwoRow{$+0.0040$}{$-0.0044$}\\[\cmsTabSkip]
        \TwoRow{\PQb tagging scale factor}{(\PQb and \PQc quarks)} & \TwoRow{$+0.0002$}{$-0.0002$} & \TwoRow{$+0.0001$}{$-0.0003$} & \TwoRow{${<}0.0001$}{${<}0.0001$} & \TwoRow{${<}0.0001$}{$-0.0002$}\\ [\cmsTabSkip]
        \TwoRow{\PQb tagging scale factor}{(light-flavor quarks and gluons)} & \TwoRow{$-0.0003$}{$+0.0004$} & \TwoRow{$-0.0003$}{${<}0.0001$} & \TwoRow{$-0.0009$}{$+0.0007$} & \TwoRow{$-0.0007$}{$+0.0005$}\\ [\cmsTabSkip]
        \OneRow{Lepton efficiencies} & \TwoRow{$-0.0002$}{$+0.0002$} & \TwoRow{$-0.0001$}{$-0.0001$} & \TwoRow{$-0.0001$}{${<}0.0001$} & \TwoRow{$-0.0004$}{$+0.0001$}\\ [\cmsTabSkip]
        \OneRow{Jet energy resolution} & \TwoRow{$-0.0028$}{$-0.0029$} & \TwoRow{$-0.0069$}{$+0.0032$} & \TwoRow{$-0.0024$}{$-0.0021$} & \TwoRow{$-0.0070$}{$+0.0026$}\\ [\cmsTabSkip]
        \OneRow{Jet energy scale} & \TwoRow{$-0.0051$}{$-0.0018$} & \TwoRow{$-0.0046$}{$+0.0065$} & \TwoRow{$-0.0046$}{$+0.0011$} & \TwoRow{$-0.0062$}{$+0.0041$}\\ [\cmsTabSkip]
        \OneRow{Background template} & \OneRow{$+0.0061$} & \OneRow{$+0.0050$} & \OneRow{$+0.0139$} & \OneRow{$+0.0016$}\\[\cmsTabSkip]
        \OneRow{PDF} & \TwoRow{$+0.0008$}{$-0.0008$} & \TwoRow{$-0.0008$}{$+0.0006$} & \TwoRow{$+0.0003$}{$-0.0004$} & \TwoRow{$+0.0003$}{$-0.0006$}\\[\cmsTabSkip]
        \TwoRow{QCD renormalization}{and factorization} & \TwoRow{$+0.0008$}{$+0.0012$} & \TwoRow{$+0.0008$}{$-0.0002$} & \TwoRow{$+0.0013$}{$-0.0033$} & \TwoRow{$+0.0007$}{$-0.0004$}\\ [\cmsTabSkip]
        \TwoRow{Initial-state}{QCD radiation} & \TwoRow{$+0.0006$}{$-0.0004$} & \TwoRow{$-0.0005$}{$+0.0004$} & \TwoRow{$+0.0017$}{$-0.0015$} & \TwoRow{$+0.0024$}{$-0.0021$}\\ [\cmsTabSkip]
        \TwoRow{Final-state}{QCD radiation} & \TwoRow{$-0.0001$}{$-0.0008$} & \TwoRow{$-0.0215$}{$+0.0122$} & \TwoRow{$+0.0053$}{$-0.0017$} & \TwoRow{$-0.0129$}{$+0.0060$}\\ [\cmsTabSkip]
        \OneRow{Color reconnection} & \TwoRow{$-0.0162$}{${<}0.0001$} & \TwoRow{$+0.0186$}{$-0.0206$} & \TwoRow{$+0.0091$}{$-0.0464$} & \TwoRow{$+0.0384$}{$+0.0304$}\\ [\cmsTabSkip]
        \OneRow{ME-PS matching} & \TwoRow{$-0.0235$}{$+0.0399$} & \TwoRow{$-0.0043$}{$+0.0177$} & \TwoRow{$-0.0185$}{$+0.0139$} & \TwoRow{$+0.0352$}{$+0.0376$}\\ [\cmsTabSkip]
        \OneRow{Underlying event} & \TwoRow{$-0.0515$}{$-0.0099$} & \TwoRow{$-0.0576$}{$+0.0355$} & \TwoRow{$-0.0082$}{$+0.0218$} & \TwoRow{$+0.0116$}{$+0.0424$}\\ [\cmsTabSkip]
        \OneRow{Flavor response} & \TwoRow{$-0.0017$}{$-0.0024$} & \TwoRow{$-0.0007$}{$+0.0024$} & \TwoRow{$-0.0033$}{$-0.0004$} & \TwoRow{$-0.0105$}{$+0.0070$}\\ [\cmsTabSkip]
        \TwoRow{Top quark}{mass variation} & \TwoRow{$+0.0049$}{$-0.0179$} & \TwoRow{$+0.0152$}{$-0.0118$} & \TwoRow{$+0.0119$}{$-0.0097$} & \TwoRow{$+0.0082$}{$-0.0046$}\\ [\cmsTabSkip]
        \OneRow{Per-event resolution} & \TwoRow{$-0.0027$}{$-0.0004$} & \TwoRow{$-0.0022$}{$+0.0040$} & \TwoRow{$+0.0023$}{$+0.0014$} & \TwoRow{$-0.0005$}{$+0.0048$}\\ [\cmsTabSkip]
        \OneRow{\WHF fraction} & \OneRow{$-0.0174$} & \OneRow{$-0.0132$} & \OneRow{$-0.0102$} & \OneRow{$-0.0098$}\\ [\cmsTabSkip]
        \TwoRow{No top quark \PT}{reweighting} & \OneRow{$-0.0008$} & \OneRow{$-0.0005$} & \OneRow{${<}0.0001$} & \OneRow{${<}0.0001$}
    \end{tabular}
\end{table}

\subsubsection{Other experimental systematic uncertainties}
The systematic uncertainty due to the modeling of pileup is estimated by shifting the total inelastic cross section up and down by 4.6\%~\cite{Sim:pileup}.
The contribution to the overall uncertainty in \Acpprime is less than 0.005\%.

To assess the uncertainty coming from the \PQb tagging scale factors, the factors are varied according to their uncertainties.
The effect of changing the heavy-flavor quark (\PQb and \PQc), and light-flavor quark and gluon (\PQu, \PQd, \PQs, and \Pg) scale factors, are calculated separately.
The two variations are combined in quadrature to give the total \PQb tagging uncertainty, which contributes $<$0.001\% to the final \Acpprime measurements.

{\tolerance=800
The uncertainties from lepton identification, isolation, and trigger efficiencies are determined by changing the corresponding scale factors according to their uncertainties.
Among the sources of experimental uncertainty, those associated with these sources have the smallest values, with a contribution of $<$0.0005\% to the total systematic uncertainty.
\par}

The jet energy scale (JES) and jet energy resolution are changed according to their {\PT}- and $\eta$-dependent uncertainties~\cite{CMS:2016lmd}.
Both impact the \Mlb distribution and contribute $<$0.007\% uncertainties to the final results.

In this paper, a background template derived from the background-enriched events in data is used.
However, this template is not identical to the predicted MC background in signal events.
The difference is considered as one of our systematic uncertainties, obtained by replacing the nominal template by the simulated one.
The resulting $\approx$0.01\% uncertainty in \Acpprime is the largest experimental uncertainty.

\subsubsection{Theoretical systematic uncertainties}
The uncertainty from the PDF used in the simulation of the \ttbar process is estimated by reweighting the corresponding variations in the NNPDF 3.1 sets~\cite{CMS:symmhessian}.
This results in one of the smallest contributions among all the sources of theoretical uncertainty with a value of $<$0.001\% in the final \Acpprime measurements.

The impact of the QCD renormalization and factorization scales on the \ttbar simulation is obtained by changing them independently during the production of the simulated samples by a factor of 0.5 or 2.
The two contributions where one scale is moved up while the other is changed down are excluded.
The total uncertainty is estimated by taking the maximum deviation from the nominal result.
The resulting uncertainties are $<$0.003\% to the final results.

The uncertainty from the modeling of the PS is obtained by changing the renormalization scale for initial- and final-state QCD radiation (ISR and FSR) up and down by a factor of 2 (for ISR) and $\sqrt{2}$ (for FSR).
The resulting uncertainties are around 0.002 and 0.02\% for ISR and FSR, respectively.

The default MC simulation uses the multiple-parton interaction (MPI) scheme for color reconnection (CR) with early-resonance decays switched off in the \PYTHIA package.
The uncertainty from this method is estimated using two other CR models within \PYTHIA, a gluon-move scheme and a QCD-inspired scheme~\cite{CMS:CR1,CMS:CR2}.
The resulting systematic uncertainty associated with CR is $<$0.05\%.

The uncertainty in the matching scale between the ME and PS is derived by varying a damping parameter in \POWHEG.
Its nominal value in simulation of $1.379\Mt$ is changed to $2.305\Mt$ and $0.8738\Mt$~\cite{Sim:CP5}.
The resulting estimation of the systematic uncertainty in \Acpprime is $<$0.04\%.

The uncertainty from modeling of the UE is estimated by varying the CP5 tune in the \ttbar MC samples~\cite{Sim:CP5}.
Among the theoretical uncertainties, this has the largest value of $\approx$0.06\% in the \Acpprime measurement.

The uncertainty coming from the jet response to gluons and \PQc, \PQb, and light quarks is estimated by varying separately the JES responses for each of the four jet flavors within their uncertainties.
A systematic uncertainty of about 0.01\% in \Acpprime was found.

The top quark mass value in the simulation is varied by $\pm 1\GeV$ to estimate the uncertainty due to this parameter, leading to a value of $\approx$0.02\%.

An average mass resolution for the reconstructed top quark and \PW boson invariant masses is used in the \chisq calculation.
The actual event-by-event resolutions depend on the detector response within different $\eta$ and $\phi$ regions.
With different detector responses, the measured mass resolutions of the reconstructed top quark and \PW boson masses change accordingly.
However, the overall effects have a negligible impact compared to using the average resolution.
To estimate the worst-case scenario, the top quark and \PW boson mass resolutions are scaled up and down by 10\% per event.
The resulting uncertainty is $<$0.005\% in the final \Acpprime measurements.

The fraction of \WHF events might be different in the background-enriched events than in the signal events because of the requirement of not having a \PQb-tagged jet.
This would cause a misestimation of the background in the signal region.
To estimate the effect of this possible bias, the \WHF events in the signal region are reweighted in the simulation by a factor of 10 in the signal region to raise the corresponding fraction.
The systematic uncertainties are estimated by replacing the original \Wjets samples, which leads to an estimated systematic uncertainty of about 0.02\%.

The measured \PT spectrum of top quarks in data is considerably softer than predicted by the MC simulations.
This feature has been seen by both the ATLAS and CMS experiments in previous measurements~\cite{CMS:2016oae,ATLAS:2019hxz,Kidonakis:2012rm,Czakon:2015owf,Czakon:2017wor,Catani:2019hip}. 
To correct for this, the simulated \PT spectrum in the nominal \ttbar samples is reweighted to match the measured distribution in data.
To estimate the uncertainty in the \Acpprime measurement from this effect, the \PT reweighting is removed.
A resulting uncertainty of $<$0.001\% is determined.

From Table~\ref{tab:acp_uncertainties}, we see that the dominant sources of systematic uncertainty are from the ME-PS matching, the UE simulation, and the correction for the \WHF content.

\section{Results}\label{sec:results}
\subsection{Asymmetry measurements}
\begin{figure}[!t]
    \centering
    \includegraphics[width=0.7\textwidth]{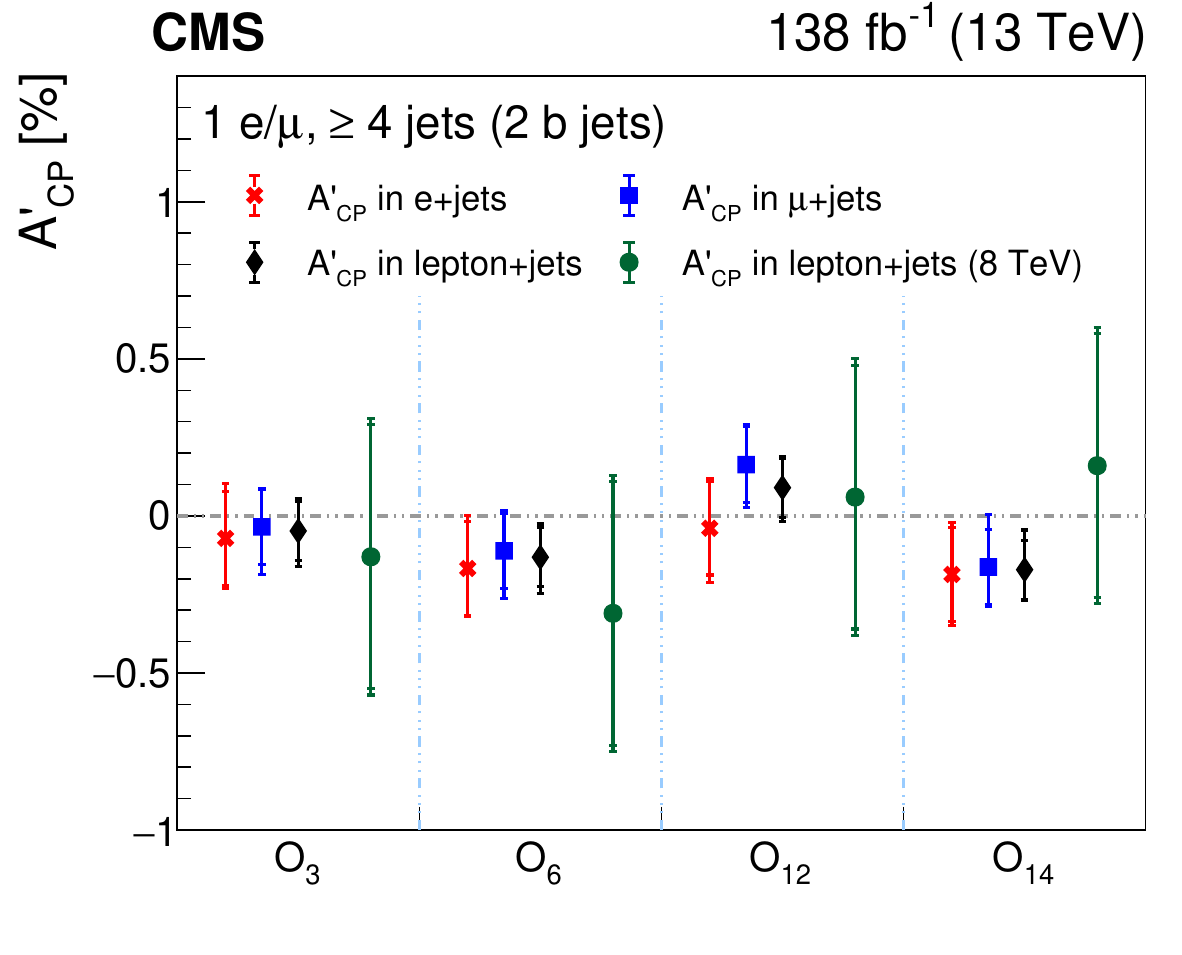}
    \caption
    {
        The measured effective asymmetries \Acpprime for each CP observable from the electron, muon, and combined lepton+jets channels.
        The results from the previous CMS measurement at \oldTeV~\cite{CPVtop:CMSresult} are shown by the green circles.
        The inner vertical bars on the symbols represent the statistical uncertainty, and the outer bars the statistical and systematic uncertainties added in quadrature.
    }
    \label{fig:unblind_result}
\end{figure}

The effective asymmetries \Acpprime are obtained after implementing the fitting procedure described in Section~\ref{sec:fitresult}.
The final results for \Acpprime in each lepton-flavor channel are displayed in Fig.~\ref{fig:unblind_result} and the values given in Table~\ref{tab:final_result}.
The correlations between the dominant sources of systematic uncertainty are around 0.3 and only affect the measurement of \Acpprime on the order of 0.001\%.
The results from the previous CMS search at \oldTeV~\cite{CPVtop:CMSresult} are shown for comparison in Fig.~\ref{fig:unblind_result}.
There is no statistically significant evidence for CPV from the present analysis in either lepton-flavor channel for any of the CP observables.
The measured \Acpprime values are in agreement with the SM expectations and have uncertainties roughly a factor of 3 smaller than the previous CMS result.

\begin{table}[!t]
    \topcaption
    {
        The measured effective asymmetries \Acpprime in percent for each of the CP observables for the electron, muon, and combined data samples.
        The first uncertainty is statistical and the second is systematic.
        The statistical uncertainties are the same for each lepton type because the numbers of signal events are the same.
    }
    \label{tab:final_result}
    \centering\renewcommand\arraystretch{1.4}
    \begin{tabular}{cccc}
        \multicolumn{4}{c}{\Acpprime (\%)}\\
        & \ejets & \mjets & Combined\\
        \hline
        \Othree
        & $-0.07 \pm 0.15\,^{+0.09}_{-0.06}$
        & $-0.04 \pm 0.12\,^{+0.02}_{-0.09}$
        & $-0.05 \pm 0.09\,^{+0.04}_{-0.07}$\\
        \Osix
        & $-0.17 \pm 0.15\,^{+0.08}_{-0.04}$
        & $-0.11 \pm 0.12\,^{+0.04}_{-0.09}$
        & $-0.13 \pm 0.09\,^{+0.05}_{-0.07}$\\
        \Otwelve
        & $-0.04 \pm 0.15\,^{+0.06}_{-0.09}$
        & $+0.16 \pm 0.12\,^{+0.04}_{-0.07}$
        & $+0.09 \pm 0.09\,^{+0.03}_{-0.05}$\\
        \Ofourteen
        & $-0.19 \pm 0.15\,^{+0.08}_{-0.07}$
        & $-0.16 \pm 0.12\,^{+0.12}_{-0.03}$
        & $-0.17 \pm 0.09\,^{+0.09}_{-0.02}$\\
    \end{tabular}
\end{table}

\subsection{Constraint on the dimensionless CEDM}
Each of the measured \Acpprime values from the combined lepton+jets channel is divided by the dilution factor for that CP observable to obtain the corrected asymmetry \Acp.
These values are independent of any detector or reconstruction bias and can therefore be used to search for CPV in the top quark sector.
The constraints on the dimensionless CEDM \dtG are determined based on the \Acp value of each CP observable, and are then combined using the best linear unbiased estimator method~\cite{LYONS1988110}, taking into account the correlation of the \dtG measurements among the CP observables.
Most of the correlation coefficient values between the CP observables are negligible, except for the one between \Othree and \Osix, which is around 0.44.
Since there is no statistically significant evidence of CPV for any CP observable, the expected dilution factor \dilution from the SM simulation given in Table~\ref{tab:dilution_factor} is used to convert the measured \Acpprime for each CP observable to the corresponding \Acp value.
Since the unbiased estimator method only uses symmetric uncertainties, the larger of the plus and minus uncertainties in \Acpprime is taken in each case.

Dedicated CP-violating \ttbar samples using the modified top quark production vertex described by Eq.~\eqref{eq:lagrangian} are simulated with different \dtG values at the generator level with \MADGRAPH to determine the relationship between \Acp and \dtG.
This relationship is parametrized by the function~\cite{CPVtop:13TeVRef}
\begin{linenomath}\begin{equation}
    \Acp = \frac{\dtG+a}{b \dtG^2+ c \dtG + d},
\end{equation}\end{linenomath}
where the parameters $a$, $b$, $c$, and $d$ are taken from a \chisq fit to the \Acp and \dtG values obtained from the different CP-violating samples.

From this relation, the dimensionless CEDM parameter \dtG is obtained from the measured \Acp value, and the uncertainty in \dtG is calculated using the full covariance matrix from the fit:
\begin{linenomath}\begin{equation}
\DsqdtG=\delvec^{T} \begin{pmatrix}
    \DsqAcp & 0 & 0 & 0 & 0 \\
    0 & \Dsqa & \cov(a,b) & \cov(a,c) & \cov(a,d) \\
    0 & \cov(b,a) & \Dsqb & \cov(b,c) & \cov(b,d) \\
    0 & \cov(c,a) & \cov(c,b) & \Dsqc & \cov(c,d)\\
    0 & \cov(d,a) & \cov(d,b) & \cov(d,c) & \Dsqd\\
\end{pmatrix} \delvec,
\end{equation}\end{linenomath}
where \DsqdtG, \DsqAcp, \Dsqa, \Dsqb, \Dsqc, and \Dsqd are the variances in \dtG, \Acp, $a$, $b$, $c$, and $d$, respectively.

\begin{figure}[!th]
    \centering
    \includegraphics[width=0.8\textwidth]{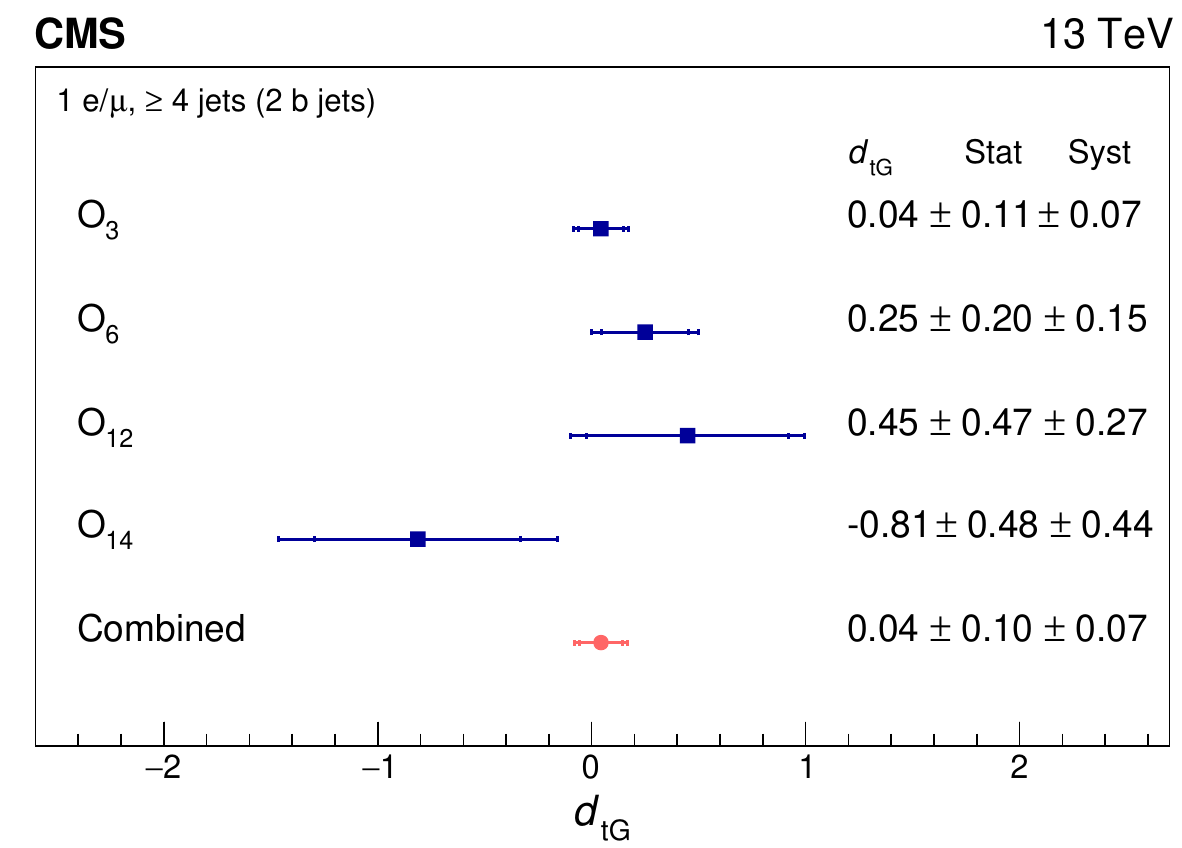}
    \caption
    {
        The measured dimensionless CEDM \dtG for each CP observable (blue squares) in the lepton+jets channel and the combined result (red point).
        The inner horizontal bars on the points represent the statistical uncertainty and the outer bar the combined statistical and systematic uncertainties added in quadrature.
    }
    \label{fig:upper_limit}
\end{figure}

The resulting CP-violating asymmetry \Acp and the dimensionless CEDM \dtG, with their statistical and systematic uncertainties, are shown in Table~\ref{tab:upper_limit} for each of the CP observables.
The final systematic uncertainties are taken as the larger value of the total systematic uncertainties between the up and down directions.
The resulting constraints on \dtG are displayed in Fig.~\ref{fig:upper_limit}.
Combining the results from the four asymmetries, leads to an overall value of $\dtG = 0.04\pm 0.10\stat\pm 0.07\syst$.
This corresponds to a limit of $\abs{\dtG} < 0.25$ at the 95\% confidence level (\CL).
The final results are in agreement with the SM expectation.

\begin{table}[!ht]
    \topcaption
    {
        The measured \Acp and corresponding \dtG values for each of the CP observables using the SM simulation predictions for the dilution factor \dilution in the combined lepton+jets channel.
        The first uncertainty is statistical and the second is systematic.
    }
    \label{tab:upper_limit}
    \centering\renewcommand\arraystretch{1.2}
    \begin{tabular}{ccc}
        CP observable & \Acp (\%) & \dtG\\
        \hline
        \Othree & $-0.10\pm0.20\pm0.14$ & $+0.04\pm0.11\pm0.07$\\
        \Osix & $-0.30\pm0.21\pm0.16$ & $+0.25\pm0.20\pm0.15$\\
        \Otwelve & $+0.12\pm0.13\pm0.07$ & $+0.45\pm0.47\pm0.27$\\
        \Ofourteen & $-0.29\pm0.16\pm0.14$ & $-0.81\pm0.48\pm0.44$\\
    \end{tabular}
\end{table}

Constraints on the CEDM of the top quark have also been set by CMS from measurements of the spin correlations~\cite{CPVtop:spin_correlation} in dilepton \ttbar events using \dt, which is the dimensionless parameter of CEDM in the convention of Ref.~\cite{Bernreuther:2013aga}.
The relation between \dt and \dtg can be expressed as $\dt=\Mt \dtg$~\cite{CPVtop:spin_correlation_dtg}.
From Eq.~\eqref{eq:cedm_conversion}, with the BSM scale $\Lambda$ at 1\TeV, the combined \dtG values from this analysis can be converted into a 95\% \CL limit of $\abs{\dt} < 0.015$.
This is competitive with the 95\% \CL limits obtained by the spin-correlation measurement~\cite{CPVtop:spin_correlation} of $-0.020 < \dt < 0.012$.

\section{Summary}\label{sec:summary}
The results of a search have been presented for the combined charge conjugate and parity (CP) violation effects in top quark-antiquark events performed in the electron+jets and muon+jets final states.
The top quark and antiquark are each assumed to decay into a bottom quark and a \PW boson, with one \PW boson decaying hadronically and the other leptonically into an electron or muon and accompanying neutrino.
This study uses data collected by the CMS experiment at the LHC from proton-proton collisions at \newTeV, corresponding to an integrated luminosity of 138\fbinv.
The CP-violating asymmetries are obtained with four different triple-product T-odd observables, where T is the time-reversal operator, constructed using linearly independent four-momentum vectors associated with the final-state particles.
The asymmetries are computed using the fitted signal.
There are no statistically significant indications of CP violation, with all the CP asymmetries being consistent with the standard model expectations.
The resulting combined measurement of the dimensionless chromoelectric dipole moment gives $\dtG=0.04\pm0.10\stat\pm0.07\syst$ and exhibits no evidence for CP-violating effects, consistent with expectations from the standard model.
These results are compatible with a previous study performed by CMS in data from \pp collisions at \oldTeV~\cite{CPVtop:CMSresult} with uncertainties improved by roughly a factor of 3.

\begin{acknowledgments}
    We are grateful to German Valencia and Alper Hayreter for providing the CEDM samples and aid in their use.

    We congratulate our colleagues in the CERN accelerator departments for the excellent performance of the LHC and thank the technical and administrative staffs at CERN and at other CMS institutes for their contributions to the success of the CMS effort. In addition, we gratefully acknowledge the computing centers and personnel of the Worldwide LHC Computing Grid and other centers for delivering so effectively the computing infrastructure essential to our analyses. Finally, we acknowledge the enduring support for the construction and operation of the LHC, the CMS detector, and the supporting computing infrastructure provided by the following funding agencies: BMBWF and FWF (Austria); FNRS and FWO (Belgium); CNPq, CAPES, FAPERJ, FAPERGS, and FAPESP (Brazil); MES and BNSF (Bulgaria); CERN; CAS, MoST, and NSFC (China); MINCIENCIAS (Colombia); MSES and CSF (Croatia); RIF (Cyprus); SENESCYT (Ecuador); MoER, ERC PUT and ERDF (Estonia); Academy of Finland, MEC, and HIP (Finland); CEA and CNRS/IN2P3 (France); BMBF, DFG, and HGF (Germany); GSRI (Greece); NKFIH (Hungary); DAE and DST (India); IPM (Iran); SFI (Ireland); INFN (Italy); MSIP and NRF (Republic of Korea); MES (Latvia); LAS (Lithuania); MOE and UM (Malaysia); BUAP, CINVESTAV, CONACYT, LNS, SEP, and UASLP-FAI (Mexico); MOS (Montenegro); MBIE (New Zealand); PAEC (Pakistan); MES and NSC (Poland); FCT (Portugal); MESTD (Serbia); MCIN/AEI and PCTI (Spain); MOSTR (Sri Lanka); Swiss Funding Agencies (Switzerland); MST (Taipei); MHESI and NSTDA (Thailand); TUBITAK and TENMAK (Turkey); NASU (Ukraine); STFC (United Kingdom); DOE and NSF (USA).
        
    \hyphenation{Rachada-pisek} Individuals have received support from the Marie-Curie program and the European Research Council and Horizon 2020 Grant, contract Nos.\ 675440, 724704, 752730, 758316, 765710, 824093, 884104, and COST Action CA16108 (European Union); the Leventis Foundation; the Alfred P.\ Sloan Foundation; the Alexander von Humboldt Foundation; the Belgian Federal Science Policy Office; the Fonds pour la Formation \`a la Recherche dans l'Industrie et dans l'Agriculture (FRIA-Belgium); the Agentschap voor Innovatie door Wetenschap en Technologie (IWT-Belgium); the F.R.S.-FNRS and FWO (Belgium) under the ``Excellence of Science -- EOS" -- be.h project n.\ 30820817; the Beijing Municipal Science \& Technology Commission, No. Z191100007219010; the Ministry of Education, Youth and Sports (MEYS) of the Czech Republic; the Hellenic Foundation for Research and Innovation (HFRI), Project Number 2288 (Greece); the Deutsche Forschungsgemeinschaft (DFG), under Germany's Excellence Strategy -- EXC 2121 ``Quantum Universe" -- 390833306, and under project number 400140256 - GRK2497; the Hungarian Academy of Sciences, the New National Excellence Program - \'UNKP, the NKFIH research grants K 124845, K 124850, K 128713, K 128786, K 129058, K 131991, K 133046, K 138136, K 143460, K 143477, 2020-2.2.1-ED-2021-00181, and TKP2021-NKTA-64 (Hungary); the Council of Science and Industrial Research, India; the Latvian Council of Science; the Ministry of Education and Science, project no. 2022/WK/14, and the National Science Center, contracts Opus 2021/41/B/ST2/01369 and 2021/43/B/ST2/01552 (Poland); the Funda\c{c}\~ao para a Ci\^encia e a Tecnologia, grant CEECIND/01334/2018 (Portugal); the National Priorities Research Program by Qatar National Research Fund; MCIN/AEI/10.13039/501100011033, ERDF ``a way of making Europe", and the Programa Estatal de Fomento de la Investigaci{\'o}n Cient{\'i}fica y T{\'e}cnica de Excelencia Mar\'{\i}a de Maeztu, grant MDM-2017-0765 and Programa Severo Ochoa del Principado de Asturias (Spain); the Chulalongkorn Academic into Its 2nd Century Project Advancement Project, and the National Science, Research and Innovation Fund via the Program Management Unit for Human Resources \& Institutional Development, Research and Innovation, grant B05F650021 (Thailand); the Kavli Foundation; the Nvidia Corporation; the SuperMicro Corporation; the Welch Foundation, contract C-1845; and the Weston Havens Foundation (USA).
\end{acknowledgments}

\bibliography{auto_generated}
\cleardoublepage \appendix\section{The CMS Collaboration \label{app:collab}}\begin{sloppypar}\hyphenpenalty=5000\widowpenalty=500\clubpenalty=5000
\cmsinstitute{Yerevan Physics Institute, Yerevan, Armenia}
{\tolerance=6000
A.~Tumasyan\cmsAuthorMark{1}\cmsorcid{0009-0000-0684-6742}
\par}
\cmsinstitute{Institut f\"{u}r Hochenergiephysik, Vienna, Austria}
{\tolerance=6000
W.~Adam\cmsorcid{0000-0001-9099-4341}, J.W.~Andrejkovic, T.~Bergauer\cmsorcid{0000-0002-5786-0293}, S.~Chatterjee\cmsorcid{0000-0003-2660-0349}, K.~Damanakis\cmsorcid{0000-0001-5389-2872}, M.~Dragicevic\cmsorcid{0000-0003-1967-6783}, A.~Escalante~Del~Valle\cmsorcid{0000-0002-9702-6359}, R.~Fr\"{u}hwirth\cmsAuthorMark{2}\cmsorcid{0000-0002-0054-3369}, M.~Jeitler\cmsAuthorMark{2}\cmsorcid{0000-0002-5141-9560}, N.~Krammer\cmsorcid{0000-0002-0548-0985}, L.~Lechner\cmsorcid{0000-0002-3065-1141}, D.~Liko\cmsorcid{0000-0002-3380-473X}, I.~Mikulec\cmsorcid{0000-0003-0385-2746}, P.~Paulitsch, F.M.~Pitters, J.~Schieck\cmsAuthorMark{2}\cmsorcid{0000-0002-1058-8093}, R.~Sch\"{o}fbeck\cmsorcid{0000-0002-2332-8784}, D.~Schwarz\cmsorcid{0000-0002-3821-7331}, S.~Templ\cmsorcid{0000-0003-3137-5692}, W.~Waltenberger\cmsorcid{0000-0002-6215-7228}, C.-E.~Wulz\cmsAuthorMark{2}\cmsorcid{0000-0001-9226-5812}
% ...
\par}
\vskip\cmsinstskip
\dag:~Deceased\\
$^{1}$Also at Yerevan State University, Yerevan, Armenia\\
$^{2}$Also at TU Wien, Vienna, Austria\\
% ...
\end{sloppypar}
%%% END EDITABLE REGION %%%
% skeleton_end
\end{document}